\documentclass[12pt,a4paper]{article}
\usepackage[utf8]{inputenc}

\usepackage[a4paper,bindingoffset=0.1in,%
            left=0.9in,right=0.9in,top=1in,bottom=1in,%
            footskip=0.25in]{geometry}

\usepackage{setspace} 
\usepackage{mathptmx} 
\usepackage{xcolor}  
\usepackage{dsfont}  
\usepackage[hidelinks, bookmarksopen=true]{hyperref} 

\usepackage{tikz} 
\usetikzlibrary{positioning, shapes, arrows, calc, arrows.meta} 

\usepackage{float} 
\usepackage{graphicx} 
\usepackage{subcaption} 
\usepackage[labelfont=bf
]{caption} 
\usepackage{enumitem} 
\usepackage{amstext} 
\usepackage{amsmath}
\usepackage{array,multirow} 
\usepackage[bottom]{footmisc} 
\usepackage{tablefootnote}
\usepackage{threeparttable} 
\usepackage{url}
\usepackage{relsize} 

\usepackage[round]{natbib}
\hypersetup{colorlinks=true, citecolor=black, linkcolor=black, urlcolor=blue}

\usepackage{hyperref}
\newcolumntype{K}[1]{>{\centering\arraybackslash}p{#1}}  
\usepackage{verbatim} 

\usepackage[normalem]{ulem} 

\usepackage{authblk}  
\begin{document}

\title{\textbf{Modeling gap acceptance behavior allowing for perceptual distortions and exogenous influences}}
\date{}

\author[1]{Ankita Sharma}
\author[1]{Partha Chakroborty}
\author[1]{Pranamesh Chakraborty}
\affil[1]{Department of Civil Engineering, Indian Institute of Technology Kanpur, India}

\maketitle

\begin{abstract}
    This work on gap acceptance is based on the premise that the decision to accept/reject a gap happens in a person’s mind and therefore must be based on the perceived gap and not the measured gap. The critical gap must also exist in a person’s mind and hence, together with the perceived gap, is a latent variable. Finally, it is also proposed that the critical gap is influenced by various exogenous variables such as subject and opposing vehicle types, and perceived waiting time. Mathematical models that (i) incorporate systematic and random distortions during the perception process and (ii) account for the effect of the various influencing variables are developed. The parameters of these models are estimated for two different gap acceptance data sets using the maximum likelihood technique. The data is collected as part of this study. The estimated parameters throw valuable insights into how these influencing variables affect the critical gap. The results corroborate the initial predictions on the nature of influence these variables must exert and give strength to the gap acceptance decision-making construct proposed here. This work also proposes a methodology to estimate a measurable/observable world emulator of the latent variable critical gap. The use of the emulator critical gap provides improved estimates of derived quantities like the average waiting time of subject vehicles. Finally, studies are also conducted to show that the number of rejected gaps can work as a reasonable surrogate for the influencing variable, waiting time. 
    
    \textbf{Key words:} Gap acceptance; Critical gap; Time perception. 
\end{abstract}

\section{Introduction} \label{sec:intro}
Gap acceptance is a basic driving behavior in which a driver who wants to cross (or merge into) an opposing traffic stream looks for gaps between vehicles in that stream and accepts an adequate gap and crosses (or merges).\footnote{The term gap, as used here, is in time scale and also includes lag} Past studies on gap acceptance (for example, \citealp{Drew1967}, \citealp{ASHWORTH1968}; \citealp{Miller1971}; \citealp{Hewitt1983}; \citealp{HAGRING2000}; \citealp{wu2012}; \citealp{Krbaleketal}) model this behavior by hypothesizing the existence of a threshold that helps a driver decide whether a gap is acceptable. This threshold or the minimum acceptable gap is referred to as \textbf{critical gap}. 

In the past, different researchers have made different assumptions about the nature of the critical gap. Some, like \cite{raff} through to {\cite{mohan2018}}, have assumed the critical gap is a constant for an individual and does not vary across the driver population. Others (\citealp{blundenetal}; \citealp{SOLBERG}; \citealp{Drew1967}; \citealp{Troutbeck}; \citealp{wu2012}) have assumed that the critical gap is constant for an individual but varies according to some distribution across drivers in a population. \cite{DAGANZO19811} went a step further and assumed that the critical gap is stochastic for an individual whose parameters have a distribution across the population.  However, \cite{DAGANZO19811} could not meaningfully estimate the parameters of the model and concluded that \textquotedblleft it seems as if the problem is inherently inestimable.\textquotedblright \ While these researchers did not attempt to explain the source of the variations in the critical gap, some like \cite{pollatchek}, \cite{polusetal}, \cite{zohdy2010} and \cite{zhao2019} among others believe that critical gap depends upon waiting time while \cite{MAHMASSANI1981} assumes it to depend on number of rejected gaps. Gap acceptance behavior or critical gap was also found to be influenced by variables like traffic volume in opposing stream (\citealp{hamed1997}; \citealp{lee2018}), type of vehicles in opposing stream (\citealp{lee2018}), crossing distance (\citealp{zhao2019}), rain intensity (\citealp{zohdy2010}; \citealp{lee2018}), speed of vehicles in opposing stream and age of driver (\citealp{hamed1997}), location of vehicle in opposing stream (\citealp{chen2013}), and visual looming (\citealp{tian2022}).

One feature is, however, common in all past gap acceptance studies --- all agree that
the decision to accept or reject a gap is taken after comparing the gap with the critical gap. While the gaps can be observed and measured, the critical gap exists in a person's mind and cannot be measured -- it is a latent variable. The comparison between the gap and the critical gap has to happen in a person's mind. 

Based on an observed gap, a driver decides in his/her mind whether to accept or reject it. This decision is then acted upon by either proceeding with the maneuver or continuing to wait -- actions that are observable. However, what is not observable is the decision-making process. It appears that there are two worlds: some of the quantities are in a world that can be observed (observable world), and some quantities are in a world that is latent (person's mind). For the decision-making to happen in the person's mind, the observed gaps must transition to this latent world (mind's world) for comparisons with the latent variable critical gap. In other words, the comparisons happen between the critical gap and perceived gaps (not the measured or observed gaps). 

Various studies (for example, \citealp{caird1994}; \citealp{horswill_2005}; \citealp{Petzoldt_2014} and \citealp{schleinitz_2016}) on time to arrival of vehicles (in the present case, gaps) have found that time evaluated/perceived by the individuals is not the same as the actual or observed time and can be affected by a variety of factors. The previous studies on gap acceptance have ignored the role of perceived gaps in the acceptance process. In all the past studies (for example, see review articles by \citealp{brilon} and \citealp{amin2015review}), it is assumed that a driver can evaluate the gap size accurately. That is, the perceived gap is assumed to be equal to the observed/measured gap. A notable, but forgotten, exception is \cite{Miller1971}, where Miller admits the possibility of drivers misjudging a gap and attempts to incorporate this misjudgment through a random error term. 

In addition to the earlier studies on \textquotedblleft time to arrival\textquotedblright \ (or gaps) there are some other models in the broad area of transportation that acknowledge that errors are made while perceiving; for example, \cite{OU2018} models the car-following behavior by including the effect of perception errors in headway, \cite{DASKALAKIS2008} talks about the perception of headway at a bus stop. \cite{NIRMALE2023} uses a discrete choice framework to accommodate drivers’ perception errors in a multi-stimuli-based model of driver behavior. While these efforts are commendable in that they acknowledge the role of perceived gaps/headways, they all model \textquotedblleft misjudgments\textquotedblright\ through random error terms. Past studies ignore the fact that \textquotedblleft misjudgments\textquotedblright\ also have a systematic component or bias. Psychophysical studies on time perception (for example, see \citealp{Vierordt} or the book by \citealp{wearden}) for more than a century have established the existence of bias in time perception. In recent work, \cite{CHAKROBORTY2021} incorporates both bias and random error while modeling perceived waiting times at a toll plaza. The present work incorporates perception while modeling and estimating the critical gap from observations on the gap acceptance behavior.

Another fallout from ignoring that a difference exists between the observable world and the latent world is that all past studies erroneously consider the estimated critical gap to be an observable world threshold. Given that one can never measure the critical gap as it is a latent variable, errors introduced by ignoring the difference between the physical world and the mind's world cannot be easily captured and corrected. This work plans to address this shortcoming by suitably converting the estimated critical gap (a latent variable) to its image or emulator in the observable world. This emulator can then be used to determine engineering quantities like delay to vehicles, queue lengths, etc.

Finally, this work proposes a framework within which the impact that various exogenous influencing variables have on the critical gap can be modeled. Specifically, the impact of (i) the vehicle type the subject driver is driving, (ii) the oncoming or opposing vehicle type, and (iii) the waiting time or number of gaps rejected is studied. The estimates of the parameters of the proposed model provide valuable insight into the critical gap and the decision-making process during gap acceptance. For example, results show that the critical gap one employs while driving a more maneuverable vehicle is shorter than when driving a less maneuverable vehicle. Results also show that drivers are more risk-averse (i.e., they have a longer critical gap) when accepting gaps when the opposing vehicle is a truck than when it is, say, a smaller vehicle. Among various other interesting insights, one that stands out is that the critical gap appears to be significantly higher when the vehicle accepting the gap has just arrived versus when it has come to a stop, irrespective of the duration of the stoppage. Of course, as this waiting time becomes longer, the critical gap progressively reduces, but never at the high rate that is seen between zero and non-zero waiting times. The mathematical model with waiting time as one of the influencing variables is reasonably complex. Thus, the number of rejected gaps was tried out as a surrogate for waiting time, as it did not introduce the kind of complexity that waiting time did. The results show that the number of rejected gaps is a good surrogate for waiting time when modeling the gap acceptance behavior.

The rest of the paper is divided into six sections. The next section describes the gap acceptance behavior in greater detail and hypothesizes on why and how the type of vehicle accepting a gap, the type of approaching opposing vehicle, and waiting time may influence the critical gap. This section also delineates the role perception plays. Section \ref{sec:math_model_ga} presents the proposed mathematical models for the problem described in Section \ref{sec:prob_stat}. The model in this section is a generalized model and takes different forms depending on the assumptions one makes about the nature of the critical gap and its possible dependence on various influencing variables. In order to clearly present the different likelihood functions that would emerge from the various assumptions, Section \ref{sec:parameter_estimation} is introduced. This section also discusses issues related to the identifiability of certain parameters when using the maximum likelihood estimation (MLE) technique.  Before introducing and discussing the estimated values and other statistics in Section \ref{sec:result_dis}, a section on the data on gap acceptance collected as a part of this study at two different sites is presented in detail. Section \ref{sec:conclusion} concludes the paper by highlighting the insights into gap acceptance behavior obtained from this study.

\section{Problem statement} \label{sec:prob_stat}
Traditionally, gap acceptance behavior is modeled by comparing the measured headway or gaps in the opposing stream with a threshold (referred to as the critical gap) in order to decide which gap will be accepted and which will be rejected. The threshold value/or parameters associated with it is/are estimated from data on gap sizes and the acceptance/rejection decisions. 
 
All the past studies ignore the fact that the comparisons between gaps and the critical gap happen in one's mind, a world that is not observable. The critical gap resides only in a person's mind and, therefore, is a latent variable. Further complicating the process is the fact that the time headways or gaps in the opposing stream are never measured by the human mind, they are only perceived. So any comparisons that give rise to decisions on actions like accept or reject must arise from a comparison between the latent variable, perceived gap, and the latent variable, critical gap. The perception of time durations has been long studied. Psychology and psychophysics literature for more than a century have established that when time durations are perceived, there are both systematic distortions and random errors (\citealp{Vierordt}; \citealp{Allan1979ThePO}, \citealp{wearden}, \citealp{CHAKROBORTY2021}). In psychophysics, various laws like Weber's law (\citealp{Allan1979ThePO}), Vierordt's law (\citealp{Vierordt}) or its derivative Stevens' law (\citealp{steven1957}), that describe aspects of distortion during perception, exist. 

\begin{figure}[htbp]
    \centering 
    \includegraphics[width=0.95\linewidth]{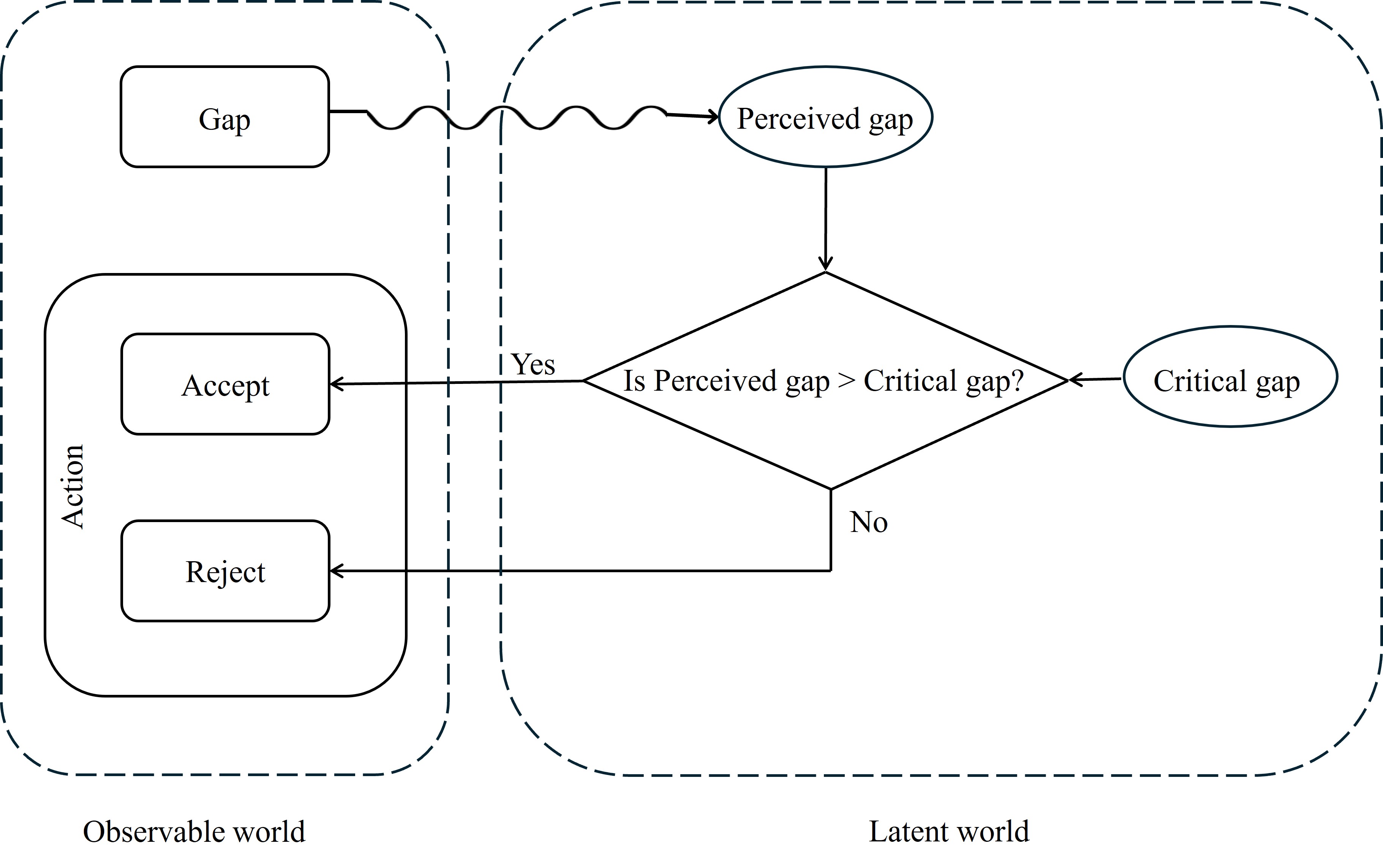}
    \caption{A representation of the gap acceptance process}
    \label{fig:gap_acc}
\end{figure}

Figure \ref{fig:gap_acc} presents a possible representation of the gap acceptance process. In the observable world, one gets to see the gaps as well as the actions like accept or reject. The gaps are perceived and compared with the latent variable critical gap and decisions on actions are taken in the mind's/latent world. In this representation, it is assumed that the decisions on the action are translated into actions without any distortion as they pass from the latent world (of the person's mind) to the observable world. However, when the gaps pass from the observable world to the latent world they go through systematic and random distortions. In the next section, a mathematical model of gap acceptance that takes into account these perceptual distortions is presented. This section also illustrates how parameters related to the distortion and parameters related to the critical gap can be estimated from observed gaps and actions. 

From an engineering perspective, however, what is required is a threshold gap value in the observable world that can be used to predict, at least statistically speaking, the actions. In the next section, a plausible strategy is devised so that one can determine a threshold value in the observable world (or an observable world critical gap) that emulates the threshold value (critical gap) in the latent world. This observable world critical gap is referred to as the emulator critical gap.

As described earlier in this section as well as in the introductory section, all existing work on gap acceptance takes, as its starting point, the existence of a threshold gap that controls accept/reject decisions as an axiomatic truth. In this work, a framework is proposed that naturally leads to the existence of such a threshold or critical gap and allows one to broaden the discussion to include features like the type of subject vehicle, type of opposing vehicle, waiting time, etc., as factors that may influence the threshold or critical gap. 

\begin{figure} [ht!]
    \centering
    \includegraphics[width=0.95\linewidth]{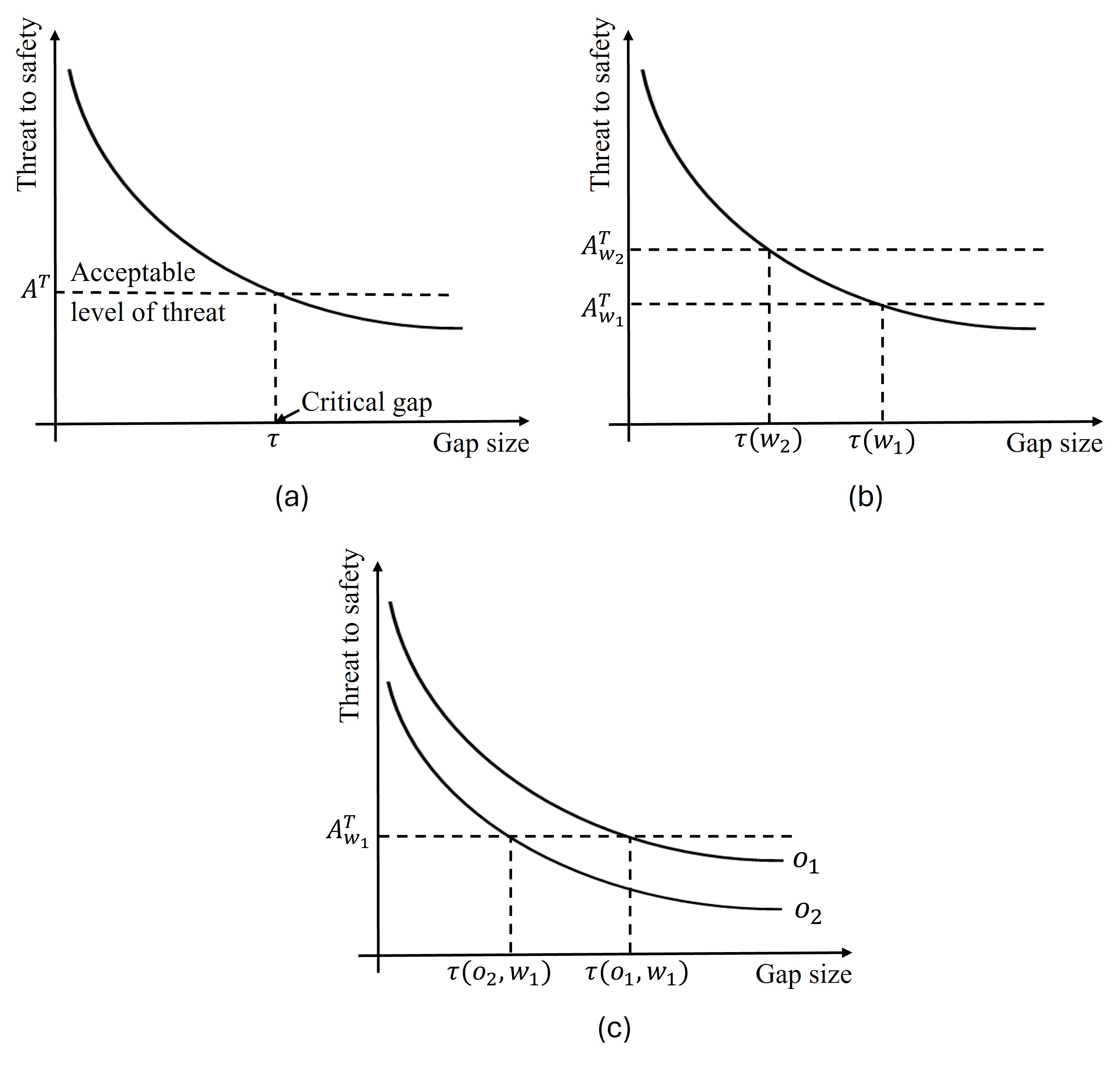}
    \caption{Variation of threat to safety with gap size.}
    \label{fig:fun_risk_ov}
\end{figure}

\citet{chakrobortyetal2004} hypothesized that a driver's actions are an outcome of the conflict between one's concern for safety and need for urgency. In the present context, the size of a gap is akin to \textquotedblleft time to collision\textquotedblright  \ and can be thought of as a determinant of the threat to safety to a driver who wishes to cross. Figure \ref{fig:fun_risk_ov}(a) provides a schematic of how the threat to safety might reduce with the size of the gap. In a stream, the number of gaps greater than a larger value will be fewer than the number greater than a smaller value. Hence, if a driver wishes to only accept gaps that are large (so as to have little threat to safety), then the driver, on average, will have to wait longer. That is, if a driver wishes to be extra safe then the driver will have to compromise on urgency. In this trade-off between safety and urgency, it is reasonable to assume that every individual has a threshold level of threat to safety $(A^T)$ above which he/she will not like to face while crossing. This assumption that drivers accept a certain level of threat while driving directly leads to the existence of a critical gap, as shown in Figure \ref{fig:fun_risk_ov}(a).  Also, as a person waits longer or rejects a large number of gaps, it can be further hypothesized that the person's sense of urgency starts dictating his/her decision-making more, and this causes the acceptable level of threat to safety to increase. That is, the person becomes more risk-taking. As shown in Figure \ref{fig:fun_risk_ov}(b), such an increase (say from $A^T_{w_1}$ to $A^T_{w_2}$ caused by waiting time increase from $w_1$ to $w_2$) implies a reduction in critical gap  $( \text{from } \tau(w_1)$ \text{to} $\tau(w_2))$. That is, one should expect that a person's critical gap reduces as waiting time increases. 

Although the size of a gap gives an idea about the time to collision, it may not completely capture the threat to safety one might face from such a collision should it happen. For example, for a given gap size, if the opposing vehicle is a two-wheeler, then the threat to safety due to a possible collision is much less than if the opposing vehicle is a heavy vehicle like a truck. Thus, different opposing vehicle types (say $o_1$ and $o_2$) must be creating different threat perceptions for the same gap size. The larger the opposing vehicle type, the higher would be the threat perception (for the same gap size) as shown in Figure \ref{fig:fun_risk_ov}(c); in the figure, $o_1$ is larger than $o_2$. The impact of such a hypothesis is that for the same acceptable threat level $(\text{say}, A^T_{w_1})$, the subject vehicle will use different critical gaps like $\tau(o_1, w_1)$ and $\tau(o_2, w_1)$ for different types of opposing vehicles. 

As the threat perception of a gap is ultimately related to the chances of a collision while crossing, then such perception must depend on the maneuverability, acceleration ability, etc., of the vehicle that wishes to cross. That is, for different types of subject vehicles, say $s_1$ and $s_2$ the threat perception for a given gap size will be different since the chances of collision will be different for different subject vehicle types. So, for different subject vehicle types, there would be different threat perception curves like the ones shown in Figure \ref{fig:fun_risk_ov}(c). This would then translate into different critical gaps for the same acceptable threat levels and opposing vehicle types. That is, $\tau(s_1,o_1,w_1)$ will be different from $\tau(s_2,o_1,w_1)$.

In this discussion, subject vehicle type, opposing vehicle type, and waiting time have been taken as instances of variables that affect or influence the critical gap. There could be other influencing variables.

In summary, the aspects of gap acceptance that are studied here are: (i) perception of gaps and decision-making during gap acceptance, (ii) the effect of subject vehicle type on the critical gap, (iii) the effect of opposing vehicle type on the critical gap, and (iv) the effect of waiting time and similar variables on the critical gap. In the next section, a mathematical model that incorporates perception as well as the effect of various influencing variables on the critical gap is developed. 

\section{Mathematical modeling of the gap acceptance process} \label{sec:math_model_ga}

Acceptance or rejection of a given gap is an outcome of the decision-making process happening in the mind's world. As described in Figure \ref{fig:gap_acc}, this process uses the perceived gap that is compared with a threshold (critical gap) to arrive at a decision. Both the perceived gap (which is a function of the observed gap) and the critical gap are latent variables. Fortunately, the decision to accept or reject, while originating in the mind's world, manifests itself as observable actions. Any mathematical model of the gap acceptance process must relate the observed gaps to the observed actions. Such a model must allow for the fact that while actions are observable, the decisions (that give rise to the actions) are made in the unobservable mind's world that works with the latent variables — perceived gaps and critical gap. 

As is apparent from Figure \ref{fig:gap_acc}, the first task in this modeling exercise will be to develop a relation between the observed gap and its perceived value. Past psychophysical studies on perceptions of time durations (note, gaps are durations of time) indicate that (i) for a given time duration, variations in perceived values increase with the duration (for example, see \citealp{Allan1979ThePO}), and (ii) smaller durations are on an average overestimated while longer durations are underestimated (\citealp{Vierordt}). Based on these primary properties of time perception and the reasonable expectation that, on average, the perceived value of a larger duration will be greater than that of a smaller duration, \citet{CHAKROBORTY2021} proposed a model to relate observed time durations to their perceived values. In this work, the same model with appropriate changes in notation is used and is presented here as Equation \ref{eq:g2gp}. 
\begin{equation} \label{eq:g2gp}
    G_p = (\alpha e^{-g/k} + \beta ) g\ \epsilon
\end{equation}

Here, $g$ is the observed size of an approaching gap, $G_p$ is the perceived gap (and is a random variable); $\alpha, \beta$ and $k $ are shape, location, and scale parameters, respectively, and  $\epsilon$ is a random error term assumed to have a log-normal distribution with cumulative distribution function (cdf) denoted as $F_\epsilon(\cdot)$, $E[\epsilon] = 1$ and $Var(\epsilon) = v$. Under these conditions, $E[G_p] = (\alpha e^{-g/k} + \beta )g$, where the term in parentheses is a multiplicative bias term. Further, as shown in \citet{CHAKROBORTY2021}, all parameters are positive and $0 < \alpha/\beta < e^2$.

The next step in the modeling process is the representation of the critical gap ($\tau$) and its comparison with the perceived gap. The simplest case is when $\tau$ is assumed to be constant and not influenced by any other variable. Later, $\tau$ is assumed to depend on one or more of the influencing variables like type of subject vehicle ($s$), type of opposing vehicle ($o$), and waiting time ($w$).

Since $\tau$ exists in the latent world (mind's world), it will be impacted by the perceived values of these influencing variables. However, it is reasonable to assume that some of these variables (for example, $s$ and $o$) remain unchanged after the perception. That is, there is no difference between their observed values and their perceived values. Of course, for a variable like $w$, such a claim is not tenable.

In order to simplify the development of the model for gap acceptance under various assumptions on $\tau$, the discussion is divided into three parts. First, the model is developed when $\tau$ is assumed to be independent of external influence. Next, this model is modified to incorporate the effect of those influencing variables, where the perceived and observed values of the variables can be assumed to be the same. Lastly, the variable waiting time, where the perceived value is different from the observed is introduced in the formulation. 

For all these cases, the model assumes population homogeneity (that is, the parameters are the same for everyone). There are two reasons for making this assumption. First, this work intends to take a first step towards incorporating perception and exogenous influencing variables in the model, and a heterogeneity assumption would have further increased the complexity of the model. Second, in the presence of perception, which causes perceived gap size to become a random quantity, the introduction of heterogeneity assumptions would have created inestimability problems like those faced by \cite{DAGANZO19811}.

\subsection{Mathematical model when \texorpdfstring{$\tau$}{tau} is independent of external influence} \label{subsec_c}
In this case, $\tau$, the critical gap in the latent (mind's) world, is assumed to be a constant. Even so, in the present approach, a given gap of size $g$ will have an associated probability of acceptance, $p_A(g)$. This is because an observed gap of size $g$ will create a random perceived gap size, $G_p$. Thus,
\begin{align} \label{eq:p_ac}
    p_A(g) &= P(G_p > \tau) \nonumber \\
           &= P\left((\alpha e^{-g/k} + \beta ) g\ \epsilon > \tau \right)\nonumber \\
           &= P \left(\epsilon > \frac{\tau}{(\alpha e^{-g/k} + \beta ) g} \right) \nonumber \\
           &=1- F_\epsilon \left(\frac{\tau}{(\alpha e^{-g/k} + \beta ) g} \right) \nonumber\\
           &=1- F_\epsilon \left(\frac{\tau/\beta}{((\alpha/\beta) e^{-g/k} + 1 ) g} \right)
\end{align}

Note, the parameters of Equation \ref{eq:p_ac} are $\tau, \alpha, \beta, k$ and $v$ where the variance $v$ is implicit in $F_\epsilon(\cdot)$. As, $F_\epsilon(\cdot)$ is the cdf of log-normal distribution with $E[\epsilon] = 1$ and $Var(\epsilon) = v$, Equation \ref{eq:p_ac} can be rewritten as, 
\begin{align} \label{eq:p_ag_c}
    p_A(g) &= 1-\Phi \left(\bigg(\ln \bigg(\frac{\tau/\beta}{\left((\alpha/\beta) e^{-g/k} + 1 \right) g}\bigg)-0.5 \ln (1+v) \bigg) 
    \!\middle/\! 
    \left(\sqrt{\ln(1+v)}\right) \right)
\end{align}

where $\Phi(\cdot)$ is the cumulative distribution function (cdf) of the standard normal distribution. 

From the above expression of $p_A(g)$, one will be able to estimate $\tau$ (or, as will be seen later, $\tau/\beta$) from observations on gap sizes and accept/reject decisions. However, from an engineering perspective, knowing the value of $\tau$ is of little consequence as it is not in the observable world. A method has to be devised to get a threshold in the observable world, say $\tau^e$ (emulator critical gap), that serves the same purpose as $\tau$ does in the latent world. That is, $\tau^e$ emulates $\tau$ in the observable world. It is this value of $\tau^e$ that can be used by engineers and planners to estimate, say, the level of service parameters like waiting time at an intersection or design length of auxiliary lanes from the length of queues of vehicles waiting to turn, etc. In all places where traditional critical gap values are used for traffic facilities design and transportation planning, one should use the emulator critical gap $\tau^e$.

In order to determine $\tau^e$, the approach taken here is as follows. Just like $\tau$ helps one decide whether a perceived gap is large enough to be accepted, $\tau^e$ should decide whether an observed gap can be accepted. Thus, if $\pi_A$ is the probability that any gap, $G$, is accepted then it can be written as,
\begin{align} \label{eq:po_c}
    \pi_A &= P(G>\tau^e) = \int_{\tau^e}^{\infty} f_G(g) \ dg \nonumber\\
           &= 1- F_G(\tau^e)
\end{align}
where $f_G(g)$ and $F_G(g)$ are respectively the known probability density function (pdf) and cumulative distribution function (cdf) of the size of gaps that arrive at the location where the gap acceptance behavior is being studied.

Now, since the latent world's decision to go on a gap of a certain size manifests itself without any distortions as accept/reject decisions in the observable world, $\pi_A$ can also be derived from $p_A(g)$ obtained in Equation \ref{eq:p_ac} as follows.
\begin{align} \label{eq:pl_c}
    \pi_A &= \int_{0}^{\infty} p_A(g)\ f_G(g) \ dg 
\end{align}

Therefore, from Equations \ref{eq:po_c} and \ref{eq:pl_c} (that provide two pathways to determine $\pi_A$) the following relation can be established.
\begin{align*} 
    1-F_G(\tau^e) &= \int_{0}^{\infty} p_A(g)\ f_G(g) \ dg\\
    &=  \mathlarger{\mathlarger{\int}}_{0}^{\infty} \left(1- F_\epsilon \left(\frac{\tau/\beta}{\left((\alpha/\beta) e^{-g/k} + 1 \right) g} \right) \right)\ f_G(g) \ dg 
\end{align*}
or, 
\begin{align}  \label{eq:tcstar_c}
    \tau^e &=  F_G^{-1} \left(  \mathlarger{\mathlarger{\int}}_{0}^{\infty} F_\epsilon \left(\frac{\tau/\beta}{((\alpha/\beta) e^{-g/k} + 1 ) g} \right) \ f_G(g) \ dg \right) 
\end{align}

Equation \ref{eq:tcstar_c} relates the emulator $\tau^e$ to the latent world's $\tau$. In the next section, the process of estimating the parameters of Equation \ref{eq:p_ac} from the data on gap sizes and accept/reject decisions is discussed. Once these parameters are estimated from the data, Equation \ref{eq:tcstar_c} is used to estimate the critical gap emulator $\tau^e$. 

This section presents the mathematical model for analyzing the gap acceptance process as introduced through Figure \ref{fig:gap_acc} and under the assumption that $\tau$ is constant and not influenced by other variables. In the next two sections, the assumption that $\tau$ is constant is relaxed. In these sections, it is assumed that $\tau$ is influenced by other variables like subject vehicle type, opposing vehicle type, etc. However, the mathematical models vary if the influencing variables are such that (i) their perceived values are the same as the observed values or (ii) their perceived and observed values differ. 
 
\subsection{Mathematical model when perceived and observed values of influencing variable are the same} \label{subsec_so}
In this section, the models of gap acceptance assume that $\tau$ is influenced by variables (like subject vehicle or opposing vehicle type) that do not go through any transformation due to perception. Let $\mathbf{y}$ be the vector of such influencing variables and $\tau(\mathbf{y})$ be the critical gap for a given value of $\mathbf{y}$. Therefore, for a given value of $\mathbf{y}$, the probability, $p_A(g,\mathbf{y})$, that a gap of size $g$ is accepted, is, 
\begin{align} \label{eq:p_ac_y}
    p_A(g,\mathbf{y}) &= P(G_p > \tau(\mathbf{y})) \nonumber \\
           &= 1-F_{\epsilon}\left( \frac{\tau(\mathbf{y})/\beta}{((\alpha/\beta) e^{-g/k} + \beta ) g} \right)
\end{align}

Again, $\tau(\mathbf{y})$ is not in the observable world, and one needs to obtain the corresponding threshold/critical gap in the observable world. Let $\tau^e(\mathbf{y})$ be the critical gap in the observable world that emulates $\tau(\mathbf{y})$ of the latent world. Like earlier (see discussion around Equation \ref{eq:po_c} through \ref{eq:tcstar_c}), since $\tau^e(\mathbf{y})$ is supposed to emulate $\tau(\mathbf{y})$, the probability that for a given $\mathbf{y}$ any gap, $G$, is accepted can be written as,
\begin{align} \label{eq:pi_A_y_lat}
    \pi_A(\mathbf{y}) &= \int_{\tau^e( \mathbf{y})}^{\infty} f_G(g) \ dg \nonumber\\
           &= 1- F_G(\tau^e( \mathbf{y}))
\end{align}

Now, like in Equation \ref{eq:pl_c}, the probability that for a given $\mathbf{y}$ any gap, $G$, is accepted, can be written as,
\begin{align} \label{eq:pi_A_y_obs}
    \pi_A(\mathbf{y}) &= \int_{0}^{\infty} p_A(g,\mathbf{y}) f_G(g) dg
\end{align}

So by equating the expressions for $\pi_A(\mathbf{y})$ in Equations \ref{eq:pi_A_y_lat} and \ref{eq:pi_A_y_obs}, $\tau^e(\mathbf{y})$ can be written as,
\begin{align} \label{eq:tcstar_y}
        \tau^e( \mathbf{y}) &=  F_G^{-1} \left(  \mathlarger{\mathlarger{\int}}_{0}^{\infty} F_\epsilon \left(\frac{\tau(\mathbf{y})/\beta}{((\alpha/\beta) e^{-g/k} + 1 ) g} \right) \ f_G(g) \ dg \right)
\end{align}

\subsection{Mathematical model when perceived and observed values of influencing variable are different} \label{subsec_w}
While variables like subject vehicle and opposing vehicle types can be incorporated in the model presented in the previous section, the variable, waiting time ($w$) cannot be a part of $\mathbf{y}$ (in the previous section) as it is not reasonable to assume that perceived waiting time value and observed waiting time value are the same. As discussed in Section \ref{sec:intro}, waiting time durations must also go through both systematic and random modifications when perceived. As in Equation \ref{eq:g2gp}, the model proposed by \citet{CHAKROBORTY2021} is used here to relate perceived waiting time, $W_p$ to $w$, the observed waiting time, 
\begin{equation} \label{eq:w2wp}
    W_p = (\alpha e^{-w/k}+ \beta )w \eta
\end{equation}
where, $\eta$ is a random error having log-normal distribution with pdf of $f_\eta(u),\ E[\eta] = 1$ and var$(\eta) = v$. Here, the parameters $\alpha,\ \beta, \ k$, and $v$ are assumed to be the same as the ones in Equation \ref{eq:g2gp}, as both the expressions represent time perception over similar time scales. The parameters could have been assumed to be different, but it was felt that the gain from the use of an additional three parameters would be insignificant.

As discussed, $\tau$ exists in the latent world, so it will be impacted by the perceived value of $w$ along with $\mathbf{y}$. In this case, let the critical gap be $\tau(\mathbf{y},W_p)$. Therefore, for a given value of $\mathbf{y}$ and $w$, the probability that a given gap of size $g$ is accepted, $p_A(g,\mathbf{y},w)$, is, 
\begin{align}
    p_A(g,\mathbf{y},w) &= P(G_p > \tau(\mathbf{y},W_p)) \nonumber \\
    &= P\left((\alpha e^{-g/k} + \beta ) g\ \epsilon > \tau(\mathbf{y},W_p) \right)\nonumber \\
    &= P\left( \epsilon > \frac{\tau(\mathbf{y},W_p)}{(\alpha e^{-g/k} + \beta ) g} \right) \nonumber
\end{align}

Assuming $\epsilon$ and $\eta$ to be independent, the above equation can be rewritten as,
\begin{align} \label{eq:pagyw}
    p_A(g,\mathbf{y},w) &= \mathlarger{\int_{0}^\infty} \left(1-F_\epsilon\left(\frac{\tau(\mathbf{y},w_p)}{(\alpha e^{-g/k} + \beta ) g}\right)  \right) f_{\eta}(u)\ du 
\end{align}

where $w_p$ is any realization of $W_p$. Note, unlike in the previous cases where the variables affecting $\tau$ were class (discrete) variables and could take a small number of values, the variable $w$, representing the waiting time, in theory, takes any non-negative real value. Hence, in this case, a function relating the variation in the critical gap with $\mathbf{y}$ and $w_p$ is as proposed.
\begin{equation} \label{eq:tau_sow}
    \tau(\mathbf{y},w_p) = a_{\mathbf{y}} e^{-w_p/l_{\mathbf{y}}} + c_{\mathbf{y}}  \qquad \qquad  w_p\ge 0
\end{equation}

where $a_{\mathbf{y}}$, $l_{\mathbf{y}}$ and $c_{\mathbf{y}}$ are positive constants that depends on $\mathbf{y}$. This particular functional form is chosen since it is expected that the critical gap will monotonically reduce with $w_p$ and will be asymptotic to some \textquotedblleft lowest value\textquotedblright \ of the critical gap. According to this function, the critical gap will monotonically fall from $a_{\mathbf{y}} + c_{\mathbf{y}}$ (when $w_p=0$) to $c_{\mathbf{y}}$ (when $w_p \rightarrow \infty$). The parameters $a_{\mathbf{y}}$ and $c_{\mathbf{y}}$ will be dependent on $\mathbf{y}$

Now, using $\tau(\mathbf{y},w_p)$ from Equation \ref{eq:tau_sow} and realizing $w_p=(\alpha e^{-w/k}+ \beta )w \ u$, Equation \ref{eq:pagyw} can be rewritten as,
\begin{align} \label{eq:p_ac_yw}
    p_A(g,\mathbf{y},w) &= \mathlarger{\int_{0}^\infty} \left(1-F_\epsilon\left(\frac{a_{\mathbf{y}} e^{-\frac{(\alpha e^{-w/k}+ \beta )w \ u}{l_{\mathbf{y}}}} + c_{\mathbf{y}}}{(\alpha e^{-g/k} + \beta ) g}\right)  \right) f_{\eta}(u)\ du \nonumber\\
    &= \mathlarger{\int_{0}^\infty} \left(1-F_\epsilon\left(\frac{\frac{a_{\mathbf{y}}}{\beta} e^{-\frac{((\alpha/\beta) e^{-w/k}+ 1 )w \ u}{l_{\mathbf{y}}/\beta}} + \frac{c_{\mathbf{y}}}{\beta}}{(\frac{\alpha}{\beta} e^{-g/k} + 1 ) g}\right)  \right) f_{\eta}(u)\ du
\end{align}

Following the development on the same lines as in Equation \ref{eq:pi_A_y_lat} through \ref{eq:tcstar_y}, the emulator critical gap in this case can be obtained as, 
\begin{align}  \label{eq:tcstar_yw}
    \tau^e( \mathbf{y},w) 
    &= F_G^{-1} \left(  \mathlarger{\mathlarger{\int}}_{0}^{\infty} \left( 1-p_A(g,\mathbf{y},w)\right) \ f_G(g) \ dg \right) 
\end{align}
which leads to the following equation that involves the evaluation of a two-dimensional integral.
\begin{align}  \label{eq:tcstar_ywn}
    \tau^e( \mathbf{y},w) 
    &=  F_G^{-1} \left(  \mathlarger{\mathlarger{\int}}_{0}^{\infty} \left( \mathlarger{\int_{0}^\infty} F_\epsilon\left(\frac{\frac{a_{\mathbf{y}}}{\beta} e^{-\frac{((\alpha/\beta) e^{-w/k}+ 1 )w \ u}{l_{\mathbf{y}}/\beta}} + \frac{c_{\mathbf{y}}}{\beta}}{(\frac{\alpha}{\beta} e^{-g/k} + 1 ) g}\right)  f_{\eta}(u)\ du\right) \ f_G(g) \ dg \right) 
\end{align}

The next section discusses how Equations \ref{eq:p_ac}, \ref{eq:p_ac_y}, and \ref{eq:p_ac_yw} are used to estimate the parameters under different assumptions related to the critical gap and its influencing variables. After obtaining the estimated parameters, Equations \ref{eq:tcstar_c}, \ref{eq:tcstar_y}, and \ref{eq:tcstar_ywn} are used to obtain $\tau^e(\cdot)$ for the different cases discussed here.

\section{Parameter estimation: preliminary discussions} \label{sec:parameter_estimation}
This section describes how the observations on gap acceptance behavior are used to estimate parameters related to time perception and critical gap. While the details of the real-world observations used to estimate the parameters are discussed in the next section and the estimates are presented in the section after that, this section outlines the format of the data in the first subsection. The second subsection develops the likelihood functions for the various cases of critical gap discussed in Sections \ref{subsec_c} through \ref{subsec_w}. In order to estimate the parameters, the likelihood functions developed here are maximized with the constraints that all parameters must be positive and $0 < \alpha/\beta <e^2$. For this purpose, the maxLik package (\citealp{maxLik}) available in R was used. Next, the estimated parameters are used to determine $\tau^e(\cdot)$. Note, the estimation process involves constraints, hence the standard errors of the parameter estimates as well as $\tau^e(\cdot)$ are obtained using the bootstrapping technique (see \citealp{efron} for more details). 

\subsection{Format of the data}
The data on gap acceptance behavior is collected using calibrated and strategically positioned video cameras at intersections. From the recordings, relevant data are extracted and stored in the format described next. 

As stated before, the vehicle at the intersection looking for a gap to accept and complete its maneuver is called the subject vehicle, and the traffic stream it moves in is called the subject vehicle stream. The vehicle at the end of an approaching gap in the opposing stream is referred to as the opposing vehicle. For every approaching gap $i$, seen by a subject vehicle, the following are noted: (i) $g_i$, the size of the gap, (ii) $o_i$, the opposing vehicle type at the end of gap $i$; $o_i \in OV$, where $OV$ is a predefined set of vehicle types in the opposing stream (iii) $s_i$, the type of subject vehicle seeing/evaluating the gap $i$; $s_i \in SV$, where $SV$ is a predefined set of vehicle types in the subject vehicle stream, (iv) $r_i$, the number of gaps rejected by the subject vehicle when it sees gap $i$ approaching, (v) $w_i$, the time the subject vehicle has already waited (at the top of the queue) when gap $i$ arrives and (vi) $\delta_i$, the decision to accept or reject taken by the subject vehicle for the gap $i$; $\delta_i = 0$ if the gap $i$ is rejected else it is 1. Note that only those situations are recorded in the data where the subject vehicle joins the queue as the first vehicle. Actions of later vehicles in the queue (if any) are not considered in the data.  
 
Typically, the types of vehicles that arrive at an intersection are not movement-dependent. That is, both the subject vehicle stream and the opposing vehicle stream have similar vehicle types, and hence, the sets $SV$ and $OV$ should be identical. However, for the problem at hand, vehicle classification is done from the perspective of driver behavior during gap acceptance, and this classification is different from those generally used to describe vehicle mix in traffic. 

Recall, it was mentioned that one expects the critical gap to be influenced by the subject vehicle type since vehicles vary in their maneuverability and acceleration ability from the stopped position. Thus, the classification strategy used for $SV$ concentrates on these features of a vehicle. It was also conjectured that the type of opposing vehicle may affect the critical gap because the threat perceived when accepting a gap might depend on the size of the approaching vehicle in the opposing stream. So while deciding on $OV$, the size of the vehicle is a major consideration for classification. Thus, since different criteria are used to decide vehicle types for $SV$ and for $OV$, these two sets are not identical. The specific details of the sets used in this work are mentioned in the section on results.

\subsection{Development of likelihood functions and determination of emulator critical gaps, \texorpdfstring{$\tau^e(\cdot)$}{taue} } \label{subsec_lik}
The expressions for the likelihoods to be used to estimate the parameters for the various $\tau(\cdot)$ models/cases introduced in the previous section are developed in this subsection. This subsection is further subdivided into five parts. The first part discusses the likelihood function when the critical gap is assumed to be constant and not influenced by exogenous variables (like subject vehicle type, $s$, type of opposing vehicle, $o$, or waiting time, $w$). This part also discusses certain issues related to the identifiability of the parameters. Part 2 develops the likelihood function when the critical gap is assumed to be only a function of $s$. The next part develops the likelihood function when the critical gap is assumed to be a function of $s$ and $o$. Part 4 presents the development of the likelihood function when the critical gap is assumed to be a function of $s$, $o$, and $w$. The inclusion of $w$, whose perceived value is not the same as its observed value, introduces certain mathematical and computational complexities. This becomes apparent from the discussion in Part 4 of the section. The next part explores the possibility of using the number of rejected gaps, $r$, as a surrogate for waiting time, $w$. The reason for exploring $r$ as a substitute for $w$ is because it is felt that the variable $r$ (a discrete quantity) may be assumed to be immutable when perceived. In each of these parts, the expression for the determination of the corresponding $\tau^e(\cdot)$ is also described.

\subsubsection{\texorpdfstring{$\tau$}{tau} independent of external influence (single valued \texorpdfstring{$\tau$}{tau})} \label{subsubsec:const_tau}
In this section, the expressions derived in the modeling section (see, Section 3) are used to develop a likelihood function for the case when $\tau$ is assumed to be independent of $s, o, \text{and } w$ (or $r$). That is, the critical gap, $\tau$, takes on a single value. 

The probability of accepting a gap of size, $g_i$, can be written from Equation \ref{eq:p_ac} as,
\begin{align} \label{eq:p_ac_i}
    p_A(g_i) &= 1-F_{\epsilon_i} \left(\frac{\tau}{(\alpha e^{-g_i/k} + \beta ) g_i}\right)=1- F_{\epsilon_i} \left(\frac{\tau/\beta}{((\alpha/\beta) e^{-g_i/k} + 1 ) g_i} \right) 
\end{align}
 
Assuming $\epsilon_i$'s to be independently and identically distributed (iid), the likelihood of getting a particular sample is, 
\begin{align} \label{eq:likelihood_const_tc}
    L(\cdot) = \prod_{i=1}^m (p_A(g_i))^{\delta_i} (1-p_A(g_i))^{(1-{\delta_i})}
\end{align}
where $m$ is the total number of gaps observed in the data, and as introduced earlier $\delta_i=1$ if the gap $i$ is accepted, else it is zero. 

Note, from Equation \ref{eq:p_ac_i}, it is apparent that only $\tau/\beta, \alpha/\beta, v \text{ (the common variance of $\epsilon_i$'s) and } k$ are estimable. That is, $\beta$ is not identifiable. Fortunately, as can be seen from Equation \ref{eq:tcstar_c},\footnote{Note, $F_\epsilon(\cdot)$ in Equation \ref{eq:tcstar_c} is same as each of the identical distributions $F_{\epsilon_i}(\cdot)$.} this non-identifiability of $\beta$ does not pose any impediment in determination of $\tau^e$. 

In each of the cases to be described in Section \ref{subsubsec_s} through \ref{subsubsec_sow}, $\beta$ remains unidentifiable for the same reason as explained in the previous paragraph. As here, this identifiability issue with $\beta$ does not affect the determination of $\tau^e(\cdot)$ in any of the cases described next.  

\subsubsection{\texorpdfstring{$\tau$}{tau} depends only on \texorpdfstring{$s$}{s} (multi-valued \texorpdfstring{$\tau$}{tau})} \label{subsubsec_s}

When the critical gap is assumed to depend only on the subject vehicle type, which is assumed to be a variable that does not go through any transformation during perception, then the influencing vector $\mathbf{y}$ (see Section \ref{sec:math_model_ga}) has one element that is $s$. So for a given $s$, the critical gap is $\tau (s) $ where $s\in SV$. Note, for this case, the critical gap takes different discrete values for different $s$; specifically, $\tau$ takes as many values as the cardinality of $SV$, $|SV|$. Therefore, from Equation \ref{eq:p_ac_y} the probability that gap $i$ of size $g_i$ is accepted by subject vehicle of type $s$, can be written as,
\begin{align}
    p_A(g_i,s) &=  1-\sum_{s\in SV} F_{\epsilon_i}\left( \frac{\tau (s)/\beta}{((\alpha/\beta) e^{-g_i/k} + 1 ) g_i} \right) . \chi_i^s
\end{align} 

where $\chi_i^s=1$ if the type of subject vehicle that accepted or rejected gap $i$ $(s_i)$ is $s$, else it is zero. 
Again, assuming the $\epsilon_i$ to be iid, the likelihood of getting a particular sample is, 
\begin{align} \label{eq:likelihood_tc_s}
    L(\cdot) = \prod_{i=1}^m \left(p_A(g_i,s)\right)^{\delta_i} \left(1-p_A(g_i,s)\right)^{(1-{\delta_i})}
\end{align}

The likelihood in Equation \ref{eq:likelihood_tc_s} is used to determine the estimates of $\tau(s)/\beta,\ \alpha/\beta,\ k$ and $v$. The emulator critical gap for each $s$, $\tau^e(s)$ can be obtained using these estimates in Equation \ref{eq:tcstar_y} and by noting that $F_\epsilon(\cdot)$ is the same distribution as the identical distributions $F_{\epsilon_i}(\cdot)$'s.

\subsubsection{\texorpdfstring{$\tau$}{tau} depends on \texorpdfstring{$s$}{s} and \texorpdfstring{$o$}{o} (multi-valued \texorpdfstring{$\tau$}{tau})}  \label{subsubsec_so}
Critical gap, as was hypothesized earlier, may also depend on opposing vehicle type, $o$.\footnote{Opposing vehicle type, $o$, is a variable that is assumed to remain unchanged during perception.} This section develops the likelihood function assuming that $\tau$ depends on $s$ and $o$. In the notation of Section \ref{sec:math_model_ga}, the influencing vector $\mathbf{y}$ now has two elements, $s$ and $o$. Since both these variables are discrete, $s \in SV$ and $o \in OV$, $\tau(s,o)$ is a multi-valued (discrete) quantity.

Therefore, from Equation \ref{eq:p_ac_y} the probability that gap $i$ of size $g_i$ having an opposing vehicle of type $o$ is accepted by a subject vehicle of type $s$ can be written as,
\begin{align} \label{eq:pa_so}
    p_A(g_i,s,o) &=  1-\sum_{\substack{s\in SV,\\ o \in OV}}F_{\epsilon_i}\left( \frac{\tau(s,o)/\beta}{((\alpha/\beta) e^{-g_i/k} + 1 ) g_i} \right) .\ \chi_i^{s,o}
\end{align} 
where $\chi_i^{s,o}=1$ if the type of subject vehicle that accepted or rejected gap $i$ $(s_i)$ is of type $s$ and the type of vehicle at the end of gap $i$ $(o_i)$ is $o$, else it is zero.
The likelihood function will be the same as in Equation \ref{eq:likelihood_tc_s} with $p_A(g_i,s,o)$ replacing $p_A(g_i,s)$. By maximizing the likelihood function, one can estimate the parameters $\frac{\tau(s,o)}{\beta}, \ \frac{\alpha}{\beta}, \ k$, and $v$. Here $ \tau^e(s,o)$ can be determined using Equation \ref{eq:tcstar_y} for each combination of $s$ and $o$ using the estimated parameters.

\subsubsection{\texorpdfstring{$\tau$}{tau} depends on \texorpdfstring{$s$}{s}, \texorpdfstring{$o$}{o} and \texorpdfstring{$w$}{w} } \label{subsubsec_sow}
As argued in Section \ref{sec:prob_stat}, the critical gap may also be influenced by waiting time, $w$,\footnote{Recall, waiting time, $w$, is a type of variable where the assumption of immutability during perception is not tenable} along with $s$ and $o$. So, from Equation \ref{eq:p_ac_yw} the probability that gap $i$ of size $g_i$ having an opposing vehicle of type $o$ is accepted by a subject vehicle of type $s$ when waiting time is $w_i$ can be written as,
\begin{align} 
    p_A(g_i,s,o,w_i) 
    &= \mathlarger{\mathlarger{\int}}_0^\infty \left(1-  \sum_{\substack{s\in SV,\\ o \in OV}}F_{\epsilon_i}\left( \frac{\frac{a_{s,o}}{\beta} e^{-\frac{((\alpha/\beta) e^{-w_i/k}+ 1 )w_i u}{l_{s,o}/\beta}} + \frac{c_{s,o}}{\beta}}{((\alpha/\beta) e^{-g_i/k} + 1 ) g_i} \right).\ \chi_i^{s,o}\right) f_{\eta}(u) du
\end{align}

where $\chi_i^{s,o}$ has the same meaning as earlier. The likelihood function can be obtained by using $p_A(g_i,s,o, w_i)$ in place of $p_A(g_i,s)$ in Equation  \ref{eq:likelihood_tc_s} which when maximized maximized will yield estimates of $\frac{a_{s,o}}{\beta}, \ \frac{c_{s,o}}{\beta},\ \frac{l_{s,o}}{\beta},\ \frac{\alpha}{\beta}, \ k$ and $v$. $\tau^e(s,o,w)$ can be determined for each $w$ using Equation \ref{eq:tcstar_ywn}.

In the results sections, likelihood functions and the $\tau^e(\cdot)$ expressions developed here are used to estimate the various parameters and determine the emulator critical gaps. 

In this section, the expressions for $p_A(g_i,s,o,w_i)$ and $\tau^e(s,o,w)$ had an added layer of complexity because the influencing variable, waiting time itself, goes through distortions due to perception. In order to see whether this mathematical and computational complexity can be reduced, an attempt was made to explore surrogates for waiting time that might not go through significant perceptual mutation.

One such surrogate for waiting time is the number of gaps rejected, $r$. Since in gap acceptance situations, empirical evidence shows a driver typically finds a gap within the first five or six gaps (see column C8 in Table \ref{tab:data_description}), $r$ is hardly ever more than four or five. In such a situation, one may assume $r$ to pass on from the observable world to the mind's world without any distortion. In the following part, a model of $\tau$ when it is influenced by $s,\ o$, and $r$ (used as a surrogate for $w$) is presented.

\subsubsection{\texorpdfstring{$\tau$}{tau} depends on \texorpdfstring{$s$}{s}, \texorpdfstring{$o$}{o} and \texorpdfstring{$r$}{r} } 
\label{subsubsec_sor}
As described earlier, the perceived value of the number of gaps rejected is assumed to be the same as the observed value of the number of gaps rejected. Also, the number of gaps rejected, $r$ is used here as a surrogate for waiting time. Hence, in this case, for a particular subject vehicle type $s$ and opposing vehicle type $o$, the function relating the variation in the critical gap with $r$ is assumed to be similar to Equation \ref{eq:tau_sow} and is as follows.
\begin{equation}
    \tau(s,o,r) = a_{s,o} e^{-r/l_{s,o}} + c_{s,o}  \qquad \qquad  r= 0,1,2,\cdots
\end{equation}

Since the perceived value of $r$ is assumed to be the same as its measured value, so in this case, the influencing vector $\mathbf{y}$ has three elements, and these are $s$, $o$, and $r$. Critical gap $\tau(\mathbf{y})$ now is $ \tau(s,o,r)$. 

Therefore, from Equation \ref{eq:p_ac_y} the probability that gap $i$ of size $g_i$ having an opposing vehicle of type $o$ is accepted by a subject vehicle of type $s$ when number of rejected gaps are $r_i$ can be written as,
\begin{align} \label{eq:pa_sor}
    p_A(g_i,s,o,r_i) &=  1-\sum_{\substack{s\in SV,\\ o \in OV}}F_{\epsilon_i}\left( \frac{\tau(s,o,r_i)/\beta}{((\alpha/\beta) e^{-g_i/k} + 1 ) g_i} \right) .\ \chi_i^{s,o} \nonumber \\
    &=  1-\sum_{\substack{s\in SV,\\ o \in OV}}F_{\epsilon_i}\left( \frac{\frac{a_{s,o}}{\beta} e^{-r/l_{s,o}} + \frac{c_{s,o}}{\beta}}{((\alpha/\beta) e^{-g_i/k} + 1 ) g_i} \right) .\ \chi_i^{s,o}
\end{align} 

where $\chi_i^{s,o}$ has the same meaning as explained earlier. The likelihood function can be obtained by using $p_A(g_i,s,o, r_i)$ in place of $p_A(g_i,s)$ in Equation  \ref{eq:likelihood_tc_s}. The likelihood function is then maximized to get the estimates $\frac{a_{s,o}}{\beta}, \ \frac{c_{s,o}}{\beta},\ l_{s,o},\ \frac{\alpha}{\beta}, \ k$ and $v$. Here $\tau^e(\mathbf{y})$ is expressed as $ \tau^e(s,o,r)$ and can be determined for each $r$ using Equation \ref{eq:tcstar_y}.

Before moving to the section on the results from the estimation of various model parameters, the gap acceptance data collected as a part of this study are presented in the next section.

\section{Data description} \label{sec:data}

\begin{figure}[htbp]
\begin{subfigure}{0.48\linewidth}
\includegraphics[width=\linewidth]{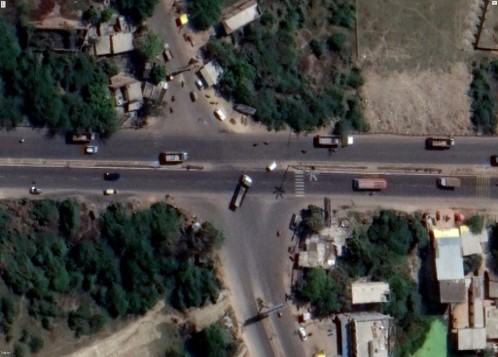}
\caption{Site-I}
\end{subfigure}
\hfill
\begin{subfigure}{0.48\linewidth}
\includegraphics[width=\linewidth]{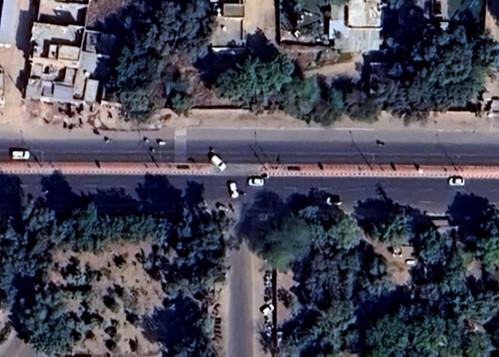}
\caption{Site-II}
\end{subfigure}%
\caption{Snapshots of data collection sites }
\label{fig:site}
\end{figure}

Sites for data collection were chosen based on the geometry of the intersection as well as the opposing volume and arrival rate of subject vehicles. Intersections that have a geometry that allows the subject vehicle ample opportunity to judge the size of the arriving gaps are ideal. The opposing volume should be such that there is a significant spread in gap sizes that a subject vehicle evaluates. The arrival rate of the subject vehicles should be moderately large so that a reasonable number of gap acceptance decisions can be observed without a queue behind the subject vehicle, as such queues may impact the gap acceptance behavior of the subject vehicle. Finally, there should be a reasonable variety in the types of opposing and subject vehicles. Based on the above criteria, two sites in India, one in Kanpur (Site-I) and the other in Bikaner (Site-II), were selected; these sites are in two different states and about 700 km apart. At each of these sites, the intersection was on a four-lane divided road. Satellite photographs of these two sites are shown in Figure \ref{fig:site}. The subject vehicles at both these intersections are those that turn right from the left-to-right movement. 

Two video cameras are used to collect the traffic data at these intersections. The cameras are placed strategically to record the complete movement of the subject vehicle as well as an adequately large part of the opposing stream. In order to be able to stitch the images from the two cameras easily, the recordings are synchronized. From the video recordings, the following data are extracted: (i) arrival times of opposing vehicles at an imaginary line drawn at an appropriate location at the intersection, (ii) opposing vehicle type,  (iii) arrival times of subject vehicles at the intersection, (iv) subject vehicle type, and (v) whether an arriving gap is rejected/accepted by the subject vehicle. 

The gap size is obtained as the difference between the arrival times of two successive vehicles\footnote{The arrival times of vehicles are arranged in ascending order irrespective of their lateral location (or lane) to find the gap size.} in the opposing stream. The waiting time for a subject vehicle, when a gap arrives, is calculated by subtracting the arrival time of the subject vehicle from the arrival time of the gap. For each gap (say, $i^{th}$ gap) the following quantities can be tabulated: (i) size of the $i^{th}$ gap, $g_i$, (ii) subject vehicle type, $s_i$ that was waiting to accept/reject the $i^{th}$ gap, (iii) opposing vehicle type, $o_i$ at the end of $i^{th}$ gap, (iv) waiting time of subject vehicle, $w_i$ when the $i^{th}$ gap arrives, and (v) number of gaps rejected, $r_i$ by the subject vehicle before the arrival of $i^{th}$ gap.

For the purpose of this analysis, the subject vehicles are classified into two types: two-wheeler, \textquotedblleft 2\textquotedblright \, and four-wheeler, \textquotedblleft 4\textquotedblright; that is, $SV =$ \{2, 4\}. Among the subject vehicles, the larger multi-axle vehicles like buses and trucks are so few that they are removed from the data before analysis. The reason for classifying subject vehicles into two-wheelers and four-wheelers is that their operational capabilities, both in terms of acceleration and maneuverability from a stopped position, are very different. The two-wheelers are typically more maneuverable. The kind of opposing vehicles is considered important in this analysis because of the threat they pose to the subject vehicle. In this analysis, they are classified into two types, namely, small, $S$ and big, $B$; that is, $OV = \{ S,B\}$. In the class $B$, all opposing vehicles of the size of passenger cars and larger are included. The rest, like motorbikes, electric rickshaws, etc., are classified as $S$. 

\begin{table}[H]
\caption{Descriptive statistics of the gap acceptance data from Sites I and II}
\label{tab:data_description}
    \makebox[1 \textwidth][c]{
    \resizebox{1 \textwidth}{!}{
    \begin{threeparttable}  
    \begin{tabular}{K{0.2cm}K{0.2cm}ccccccccc}
        \hline
        \multicolumn{2}{c}{\multirow{3}{*}{Site}} & Type of   & Tot. no. of  & Tot. no. of  & \% of gap   & No. of veh.  & 75 \%ile & 75 \%ile no. & \multicolumn{2}{c}{Gap} \\ 
        \vspace{-200 pt}
        & & subject   & subject  & rejected & with $S$ as   & accepting & waiting &  of rejected  & \multicolumn{2}{c}{size \tnote{*}} \\
        \vspace{-200 pt}
       & & vehicle & vehicle &  gaps &  opp. veh. &   first gap &  time \tnote{*} &  gaps &  Mean & SD \\

       \multicolumn{2}{c}{(C1)} & (C2)   & (C3)  & (C4)  & (C5) &(C6)  & (C7)  & (C8) &  \multicolumn{2}{c}{(C9)} \\
       
        \hline
        
        \multirow{2}{*}{\rotatebox{90}{Kanpur \ }}&\multirow{2}{*}{\rotatebox{90}{(Site-I)\ \ }}  & 2 & 463 & 1047 & 26.89 & 117 & 6.19 & 3 & \multirow{2}{*}{3.78} &  \multirow{2}{*}{4.61} \\ 
        \cline{3-9}
        && 4 & 1055 & 2808 & 25.14 & 195 & 7.90 & 4 & & \\ 
        \hline 
        \multirow{2}{*}{\rotatebox{90}{Bikaner \ }}&\multirow{2}{*}{\rotatebox{90}{(Site-II)\ \ }} &2 & 869 & 2927 & 66.78 & 162 & 6.90 & 4 & \multirow{2}{*}{2.56} & \multirow{2}{*}{3.35} \\ 
        \cline{3-9}
        & & 4 & 307 & 1088 & 65.88 & 56 & 7.86 & 5 &  &  \\ 
        \hline
    \end{tabular}
    \begin{tablenotes}
    \item[*] Waiting times and gap sizes are in seconds. 
    \end{tablenotes}
    \end{threeparttable}
}
}
\end{table}

The descriptive statistics of the data from the two sites are shown in Table \ref{tab:data_description}. Column C3 provides the total number of subject vehicles for each type that are observed. Column C4 details the total number of gaps rejected by each vehicle type. Note, since each vehicle accepts exactly one gap, the total number of gaps evaluated by each vehicle type during the gap acceptance process can be obtained by summing columns C3 and C4. Column C5 lists the percentage of the total evaluated gaps that had a small vehicle ($S$) at the end. Deducting this quantity from 100 gives the percentage with big opposing vehicles ($B$). Some vehicles accept the very first gap\footnote{Recall, in this paper, the term gap subsumes the term lag.} that they evaluate. The number of such situations is provided in Column C6. All subject vehicles that do not accept the first gap face a waiting time greater than zero, and those who accept the first gap have a zero waiting time. Column C7 lists the $75^{th}$ percentile value of the waiting times at these sites. Similarly, the $75^{th}$ percentile value of the number of gaps rejected by a subject vehicle before accepting a gap is shown in Column C8. The last column provides (C9) the mean and standard deviation of the gap sizes that arrive at each of the two sites. 

It can be seen from Table \ref{tab:data_description} that the number of observations at each site is reasonably large with ample variations in the influencing variables. The next section presents and discusses the parameter estimates obtained using the process developed in Section \ref{sec:parameter_estimation} and the data introduced in this section.
 
\section{Parameter estimates and related discussion} \label{sec:result_dis}
The parameters associated with critical gap under the various assumptions introduced in the previous sections are estimated using the data on gap acceptance collected as a part of this study. In this section, these estimates are presented. The required mathematical analyses and different likelihood functions needed for the maximum likelihood estimation are all presented in earlier sections. 

Recall, all the parameters estimated here are those of the latent variables. By themselves, they are not useful to the engineers and planners. Thus, the concept of an emulator critical gap, $\tau^e$, that is a reflection of the latent world critical gap in the observable world is created and methods to compute it under different assumptions on the critical gap are provided in relevant earlier sections. Here, the values of $\tau^e$ are also presented.

\subsection{\texorpdfstring{$\tau$}{tau} independent of external influence (single valued \texorpdfstring{$\tau$}{tau})}

In the past, the critical gap has often been considered as a constant value that is not influenced by any other measurable features of the gap acceptance environment. Therefore, in this section critical gap is estimated assuming that it is independent of any external influence. The analysis and likelihood function relevant for this section are described in Section \ref{sec:math_model_ga} and Section \ref{subsubsec:const_tau}, respectively. The parameters estimated are $\alpha/\beta,\ v ,\ k ,\ \tau/\beta$, and their estimated values are provided in Table \ref{tab:const_tau}. All parameter estimates are statistically significant.

\begin{table}[htbp]
    \centering
    \caption{Parameter estimates when $\tau$ is independent of external influence}
    \label{tab:const_tau}
    \begin{tabular}{cccccccc}
        \hline
         \multirow{2}{*}{Parameter} &\multicolumn{3}{c}{Site-I} & & \multicolumn{3}{c}{Site-II} \\
         \cline{2-4} \cline{6-8}
           & Estimate & SE& Max. $LL$ & & Estimate & SE& Max. $LL$     \\
         \hline 
           $\alpha/\beta$& 7.39 & 0.61 &\multirow{4}{*}{-1691.00} & & 6.22 & 1.45 & \multirow{4}{*}{-1397.86}  \\  
           $v$           & 0.32 & 0.03 & & & 0.32 & 0.03 &   \\
           $k$           & 0.47 & 0.05 & & & 0.32 & 0.06 &   \\
           $\tau/\beta$  & 4.33 & 0.09 &  & &3.45&0.06 &    \\      
           \hline
           \textbf{$\tau^e$}  & \textbf{4.40}  & 0.07
           &  & &\textbf{3.60}&  0.05 &  \\
         \hline
    \end{tabular}
\end{table}

Recall, analytically, the restriction $0 < \alpha/\beta < e^2$ was imposed. The estimates presented here incorporated this restriction. For Site-I, the estimated value of $\alpha/\beta = 7.39$\footnote{Note $e^2 \approx  7.39$} with a 95\% confidence interval obtained using bootstrap as $[5.84-7.39]$. A relevant question is how the estimates would change had such a restriction not been imposed. The restriction on $\alpha/\beta$ arises from the expectation that, on average, the perceived value of a larger duration will be greater than that of a smaller duration. If one were to relax this assumption, then the only restriction on the $\alpha/\beta $ would be that $\alpha/\beta > 0$. The estimation process was repeated with only $\alpha/\beta > 0$. In this case, the estimate of $\alpha/\beta $ is 7.79 with other parameter estimates remaining approximately the same. The reason for this is that numerically, there is a negligible difference between the expected perceived gaps (for a given gap) computed using the two different estimates of $\alpha/\beta$.

In order to compare the estimates of the critical gap provided in Table \ref{tab:const_tau} with those obtained from the use of existing methods, the data collected here is also used to estimate the critical gap using methods proposed by \citet{raff}, \citet{ASHWORTH1968} and \citet{Troutbeck}. 

Before discussing the results of the comparison, note: (i) none of these existing methods distinguish between the latent world and the measurable/observable world; based on the way these methods approach the determination of critical gap, their estimates are only comparable with the $\tau^e$, (ii) method proposed by \citet{ASHWORTH1968} and \citet{Troutbeck} assume critical gap to have a distribution; but as suggested by \citet{Troutbeck}, the expected value of the distribution is used as the critical gap, (iii) Troutbeck and Brilon's method requires one to remove all datasets where the accepted gap is smaller than the largest rejected gap; in that sense this method does not use all the information available in the data for estimation of critical gap, and (iv) since critical gaps can never be measured, the comparison of two computed values (one from the existing methods and $\tau^e$) is not ideal; rather a measurable quantity that is influenced by the estimated critical gap can be used for comparison. One such quantity is the average waiting time of subject vehicles. Given the estimated critical gap and the gap distribution, one can compute the expected or average waiting time, C-AWT, (details are given in Appendix \ref{app:exp_wt}). One can also determine the average of the observed waiting times, O-AWT, of the subject vehicles from the data. These can then be compared to see which estimation method of critical gap provides a more acceptable estimate of the waiting time.

Table \ref{tab:ewt_const} provides the critical gap values (in seconds) as estimated from the three existing methods discussed here and the proposed method. All the methods agree that the critical gap values at Site-II are less than those at Site-I. The table also provides (in seconds) the C-AWT from each of the four methods and O-AWT (mentioned parenthetically in the table). The proposed method gives C-AWTs that are nearest to the O-AWTs. Note, the O-AWTs for Troutbeck and Brilon's method are different from the other O-AWTs since the conditions laid out by Troutbeck and Brilon's method restrict the amount of usable data to approximately 75\% of the total data, and the O-AWTs are computed on this reduced data set.

\begin{table}[htbp]
    \centering
    \caption{Comparison of critical gaps and waiting times obtained using different methods}
    \label{tab:ewt_const}
    \begin{threeparttable}
    \begin{tabular}{c cc c cc }
        \hline
         \multirow{2}{*}{Method} &\multicolumn{2}{c}{Critical gap} & & \multicolumn{2}{c}{C-AWT (O-AWT)} \\
         \cline{2-3} \cline{5-6} 
           &  Site-I   & Site-II   && Site-I   & Site-II         \\
         \hline       \citeauthor{raff}                & 3.70  & 2.80   &&  3.42 (5.34)   &  2.94 (5.36) \\
           \citeauthor{ASHWORTH1968}           & 3.69  & 3.00   &&  3.32 (5.34)   &  3.35 (5.36)   \\
          
                   \citeauthor{Troutbeck}\tnote{*}         & 4.61  & 3.61   &&  5.58 (4.11)   &  5.29 (4.21)   \\
      \hline
      Proposed Method $(\tau^e)$                & 4.40  & 3.60   &&  4.94 (5.34)   &  5.05 (5.36)  \\
         \hline  
    \end{tabular}
    \begin{tablenotes}
    \item[*] \footnotesize Restrictions of Troutbeck's method implied only 75\% of the data could be used to estimate the critical gap using this method. The expected waiting time and the average waiting time are calculated over this reduced data set. 
    \end{tablenotes}
\end{threeparttable}

\end{table}

From the comparison of the computed average waiting times with the observed average waiting times, it appears that the proposed method's estimates of $\tau^e$ are the most acceptable. 

\subsection{\texorpdfstring{$\tau$}{tau} depends on \texorpdfstring{$s$}{s} (multi-valued \texorpdfstring{$\tau$}{tau})} \label{subsec_result_s}

Based on the hypothesis that the subject vehicle type, $s$, may influence the critical gap, the observed average waiting times (O-AWT) for two-wheelers and for four-wheelers (i.e., for $s=2$ and $s=4$) were computed separately. At each of the two sites, the O-AWTs for different $s$ turned out to be different. At Site-I, O-AWT for $s=2$ is 4.56 seconds, and for $s=4$ it is 5.68 seconds. At Site-II, O-AWT for $s=2$ is 5.12 seconds, and for $s=4$ it is 6.04 seconds. For each site, the difference is statistically significant. This lends credence to the hypothesis that the critical gap is influenced by $s$. 

To analyze the dependence, the parameters of the model introduced in Section \ref{sec:math_model_ga} are estimated by maximizing the likelihood function derived in Section \ref{subsubsec_s}. The estimated parameters, the different observed average waiting times (O-AWT) as well as the computed average waiting times (C-AWT) are presented in Table \ref{tab:tau_s}.  Recall, (i) the perception parameters are assumed to be independent of $s$, and (ii) $\tau(2) \ (\tau(4))$  is the critical gap when $s$ is a two-wheeler (four-wheeler). 

\begin{table}[htbp]
    \centering
    \caption{Parameter estimates when $\tau$ depends on $s$}
    \label{tab:tau_s}
    \makebox[1 \textwidth][c]{
    \resizebox{1 \textwidth}{!}{
    \begin{threeparttable}
    \begin{tabular}{ccccccccccc}
        \hline
         Subject& \multirow{3}{*}{Parameter}&\multicolumn{4}{c}{Site-I} & & \multicolumn{4}{c}{Site-II} \\
         \cline{3-6} \cline{8-11}
         vehicle &  & \multirow{2}{*}{Estimate} & \multirow{2}{*}{SE}  & \multirow{2}{*}{Max. $LL$}&C-AWT& & \multirow{2}{*}{Estimate} & \multirow{2}{*}{SE}& \multirow{2}{*}{Max. $LL$}&C-AWT \\
        type, $s$ &  &  &   & &(O-AWT)& &  & & & (O-AWT) \\
         \hline

          &$\alpha/\beta$  & 7.39  & 0.60 &  \multirow{5}{*}{-1683.87} &   & & 6.18 & 1.43 & \multirow{5}{*}{-1384.00} & \\ 
          &$v$             & 0.31  & 0.03 &  &   & & 0.30 & 0.03 &  & \\
          &$k$             & 0.48  & 0.05 &  &    & & 0.32 & 0.06 && \\
         2  &$\tau(2)/\beta$ & 4.02  & 0.11 & & & &3.31&0.07 &  &  \\
         4 & $\tau(4)/\beta$ & 4.47  & 0.10 &    &  & & 3.91 & 0.10 &  & \\
           \hline
         2 &  \textbf{$\tau^e(2)$}  & \textbf{4.0}  & 0.10 & &4.04 (4.56) & &\textbf{3.4}& 0.05  & &4.58 (5.12) \\
         4 &  \textbf{$\tau^e(4)$}  & \textbf{4.5}  & 0.08 & &5.18 (5.68) & &\textbf{4.0}& 0.09  & &6.66 (6.04) \\
         \hline
    \end{tabular}

    \end{threeparttable}
    }}
\end{table}

As expected, for each site, $\tau^e(2)$ is statistically significantly less than $\tau^e(4)$.\footnote{A comparison of $\tau(2)/\beta$ and $\tau(4)/\beta$ for each site yield a similar observation. Note $\tau(2)/\beta$ and $\tau(4)/\beta$ for a given site are comparable as $\beta$ is the same in both $\tau(2)/\beta$ and $\tau(4)/\beta$.} As before, the emulator critical gaps ($\tau^e$) are site specific and are shorter for Site-II. The computed waiting times for the different subject vehicle types and sites are close to the observed values. Note that the model presented in the previous section, where $\tau$ is assumed to be independent of any external influence, can be viewed as a restricted version of when $\tau$ depends on the subject vehicle type; the restriction is $\tau(2) = \tau(4) = \tau$. At each site, the likelihood ratio test suggests that the null hypothesis that the restriction is true is rejected at 0.05 level of significance. This again indicates that $s$ is a significant influencing variable and an important determinant of the critical gap.

\subsection{\texorpdfstring{$\tau$}{tau} depends on \texorpdfstring{$s$}{s} and \texorpdfstring{$o$}{o} (multi-valued \texorpdfstring{$\tau$}{tau})}
In the previous section, an analysis of average delay (waiting time) to different subject vehicle types lent credence to the hypothesis that the critical gap depends on subject vehicle types. While it is possible to determine waiting time for a given subject vehicle (and hence for a given subject vehicle type), it is not feasible to determine waiting time specific to an opposing vehicle type as any subject vehicle goes through gaps with different opposing vehicle types while waiting to cross. For this reason, one cannot simply compute waiting time specific to a particular $s$ and $o$ combination in an attempt to establish the influence of $o$ on the critical gap. Thus, in order to see whether there is any merit in the assumption that opposing vehicle type impacts critical gap, the proportion of rejected gaps for different gap sizes and for different $s$ and $o$ combinations are analyzed. The reason for studying the proportion of rejected gaps is that this fraction for a given size range of gaps depends on the critical gap and hence may throw some light on the hypothesis that opposing vehicle type is also an influencing variable.  

At each site, all gaps with a specific opposing vehicle type, $o$, that are evaluated by a specific subject vehicle type, $s$, are identified and clubbed into four bins, namely, 0--2.5 seconds, 2.5--5 seconds, 5--7.5 seconds, and greater than 7.5 seconds. Therefore, for a given $s$ and $o$ combination, namely $2S,\ 2B,\ 4S \text{ and } 4B$ at each site, there are four bins. For each bin the fraction of rejected gaps, $\rho_{so}$, is determined. As expected, the $\rho_{so}$ values for the largest bin (i.e. gaps of size greater than 7.5 seconds) are close to zero, and hence, this bin size is dropped from further analysis. At each site, for each of the remaining three bins, two comparisons are made, namely, $\rho_{2S}$ with $\rho_{2B}$ and  $\rho_{4S}$ with $\rho_{4B}$ leading to six comparisons at each site or twelve comparisons in all.\footnote{Note in each comparison of the $\rho_{so}$ values, like $\rho_{2S}$ with $\rho_{2B}$, the $s$ is held constant and only the $o$ varies.} Out of these twelve comparisons, nine yields statistically significant difference between the $\rho_{so}$ values, suggesting that the fractions of rejected gaps for a given $s$ are different for different $o$. 

The above results provide a reasonable indication that in addition to the subject vehicle type, the opposing vehicle type also has an influence on the critical gap. To analyze this dependence more rigorously, the parameters of the model introduced in Section \ref{sec:math_model_ga} are estimated using the likelihood function introduced in Section \ref{subsubsec_so}. The results are shown in Table \ref{tab:tau_so}. As the model where $\tau$ depends only on $s$  can be viewed as a restricted version of the model when $\tau$ depends on both $s$ and $o$; the likelihood ratio test was performed to compare the likelihood of these two models given in Table \ref{tab:tau_s} and \ref{tab:tau_so} for each site. The null hypothesis that the restrictions ($\tau(2,S)=\tau(2,B)=\tau(2)$ and $\tau(4,S)=\tau(4,B)=\tau(4)$) are true is rejected at 0.05 level of significance for both the sites. These results confirm the hypothesis that the opposing vehicle type is an influencing variable for the critical gap. 

Further, as expected, for each site for a given subject vehicle type the emulator critical gap for smaller opposing vehicles is statistically significantly less than that for bigger opposing vehicles (i.e., $\tau^e(2,S) < \tau^e(2,B)$ and $\tau^e(4,S) < \tau^e(4,B)$). Also, the emulator critical gap for two-wheelers is statistically significantly less than that of four-wheelers (as in Section \ref{subsec_result_s}) for each opposing vehicle type (i.e., $\tau^e(2,S) < \tau^e(4,S)$ and $\tau^e(2,B) < \tau^e(4,B)$).\footnote{Comparisons of $\tau(\cdot,S)/\beta$ and $\tau(\cdot,B)/\beta$ for each subject vehicle type and $\tau(2,\cdot)/\beta$ and $\tau(4,\cdot)/\beta$ for each opposing vehicle type at each site leads to similar observations. Also, as mentioned before, these comparisons can be made because $\beta$ (though non-identifiable) at a given site remains the same.} These observations also support the hypothesis that the threat perceived by a subject vehicle is not only a function of the gap size but also depends on the opposing vehicle type at the end of the gap and that a bigger vehicle poses a larger threat. 

\begin{table}[htbp]
    \centering
    \caption{Parameter estimates when $\tau$ depends on $s$ and $o$}
    \label{tab:tau_so}
    \begin{tabular}{cccccccc}
        \hline
         \multirow{2}{*}{Parameter} &\multicolumn{3}{c}{Site-I} & & \multicolumn{3}{c}{Site-II} \\
         \cline{2-4} \cline{6-8}
           & Estimate & SE& Max. $LL$ & & Estimate & SE& Max. $LL$     \\
         \hline
          $\alpha/\beta$      & 7.39  & 0.59 & \multirow{7}{*}{-1658.32}   & &6.04 & 1.47 & \multirow{7}{*}{-1367.90}  \\  
          $v$                 & 0.30  & 0.03 &   & &0.30 & 0.03 &   \\
          $k$                 & 0.48  & 0.05 &   & &0.32 & 0.06 &   \\
          $\tau(2,S)/\beta$  & 3.62  & 0.19 &  & &3.12& 0.08&    \\
          $\tau(2,B)/\beta$  & 4.20  & 0.13 &   & &3.76 & 0.10 &       \\
           $\tau(4,S)/\beta$  & 3.82  & 0.15 &   & &3.77 & 0.12 &       \\           
           $\tau(4,B)/\beta$  & 4.75  & 0.11 &   & &4.18 & 0.16 &        \\
           \hline
           \textbf{$\tau^e(2,S)$}  & \textbf{3.6}  & 0.18 &  & &\textbf{3.2}& 0.06  &  \\
           \textbf{$\tau^e(2,B)$}  & \textbf{4.2}  & 0.14 &  & &\textbf{3.9}& 0.09  &  \\         \textbf{$\tau^e(4,S)$}  & \textbf{3.8}  & 0.13 &  & &\textbf{3.9}& 0.11  &  \\          
           \textbf{$\tau^e(4,B)$}  & \textbf{4.8}  & 0.10 &  & &\textbf{4.3}&  0.15 &  \\
         \hline
    \end{tabular}
\end{table}

\subsection{\texorpdfstring{$\tau$}{tau} depends on \texorpdfstring{$s$}{s}, \texorpdfstring{$o$}{o} and \texorpdfstring{$w$}{w}}
\label{subsec_res_sow}
It was hypothesized in Section \ref{sec:prob_stat} that waiting time may also affect the critical gap. If this hypothesis is tenable, then one should see a difference in gap acceptance behavior between when the waiting time is small and when it is large. To see whether this can be observed from the data, the following exercise is undertaken. At each site, gaps with a specific opposing vehicle, $o$, that are evaluated by a specific subject vehicle, $s$, when waiting time, $w$, is zero are identified. The same is done for waiting time, $w$, greater than zero. Essentially, for each site, for every $s$ and $o$ combination two sub-classes are created, $w=0$ and $w>0$. Thus, there are eight combinations now, for example, $2S$ with $w=0$, $2S$ with $w>0$, $2B$ with $w=0$, $2B$ with $w>0$, and so on. For each combination, the gaps, as in the previous section, are grouped into three classes (0--2.5 seconds, 2.5--5 seconds, and 5--7.5 seconds), giving rise to 24 combinations (12 combinations for each site). For each combination, a comparison of the proportion of gaps rejected will show whether, at a given site for a given size range of gaps and $s,\ o$ combinations, there is any difference in the way drivers decide when $w=0$ and when $w>0$. In 18 of these 24 combinations/comparisons, the proportion of gaps rejected when $w=0$ is statistically different from when $w>0$. This indicates that waiting time, as envisaged, influences the critical gap. 

\begin{table}[htbp]
    \centering
    \caption{Parameter estimates when $\tau$ depends on $s$, $o$ and $w$}
    \label{tab:tau_sow}
    \begin{tabular}{cccccccc}
        \hline
         \multirow{2}{*}{Parameter} &\multicolumn{3}{c}{Site-I} & & \multicolumn{3}{c}{Site-II} \\
         \cline{2-4} \cline{6-8}
           & Estimate & SE& Max. $LL$ & & Estimate & SE& Max. $LL$     \\
         \hline
           $\alpha/\beta$  & 7.39  &  0.47 & \multirow{15}{*}{-1510.02} & &5.41 &1.72 &  \multirow{15}{*}{-1310.88} \\  
           $v$  & 0.24  & 0.02 & & & 0.26& 0.03 &   \\
           $k$  & 0.46  & 0.04 & & & 0.34& 0.07   & \\
           $a_{2,S}/\beta$  & 1.50  & 0.43 &  & &1.33&0.19 &    \\
           $c_{2,S}/\beta$  & 3.06  & 0.38 &                    & &2.84&0.09 &                      \\
           $l_{2,S}/\beta $ & 1.71  & 2.27 &                    & &0.74&0.14 &                      \\
           $a_{2,B}/\beta$  & 2.15  & 0.34 &                    & &1.59&0.31 &                       \\
           $c_{2,B}/\beta$  & 3.62  & 0.14 &                    & &3.43&0.12 &                      \\
           $l_{2,B}/\beta $ & 0.79  & 0.26 &                    & &0.53&0.17 &                      \\
           $a_{4,S}/\beta$  & 3.02  & 0.31 &                    & &1.14&0.35 &                       \\
           $c_{4,S}/\beta$  & 2.89  & 0.21 &                    & &3.52&0.17 &                      \\
           $l_{4,S}/\beta $ & 1.88  & 0.63 &                    & &0.60&0.25 &                      \\
           $a_{4,B}/\beta$  & 2.77  & 0.27 &                    & &2.80&0.71 &                      \\
           $c_{4,B}/\beta$  & 3.98  & 0.09 &                    & &3.83&0.22 &                      \\
           $l_{4,B}/\beta $ &
 1.23  & 0.23 &                    & &0.34&0.33 &                      \\

         \hline
    \end{tabular}
\end{table}
As before, to analyze this dependence further, the parameters of the model introduced in Section \ref{sec:math_model_ga} are estimated using the likelihood function derived in Section \ref{subsubsec_sow}. The estimated parameters are shown in Table \ref{tab:tau_sow}. A comparison (using the likelihood ratio test) of log-likelihood values (for each site) given in Tables \ref{tab:tau_so} and \ref{tab:tau_sow}, results in rejecting the null hypothesis that restrictions $a_{2,S}=a_{2,B}=a_{4,S}=a_{4,B}=0$ are true at 0.05 level of significance. This therefore suggests that in addition to $s$ and $o$, $w$ is a significant influencing variable in the determination of $\tau$.

Also, while proposing Equation \ref{eq:tau_sow}, it was hypothesized that the critical gap should monotonically reduce with waiting time and have some minimum positive value. For this reason, $a_\textbf{y}, c_\textbf{y}$ and $l_\textbf{y}$ (in Equation \ref{eq:tau_sow}) are constrained to be positive during estimation. Even if these constraints are removed, the estimated parameters ($a_{2,S},\ a_{2,B},\ a_{4,S},\ a_{4,B},\ c_{2,S},\ c_{2,B}$ and so on) remain positive, giving further evidence in support of the hypothesis on the behavior of critical gap with $w$. 

\begin{figure} [h!]
    \centering
    \includegraphics[width = 0.8 \linewidth]{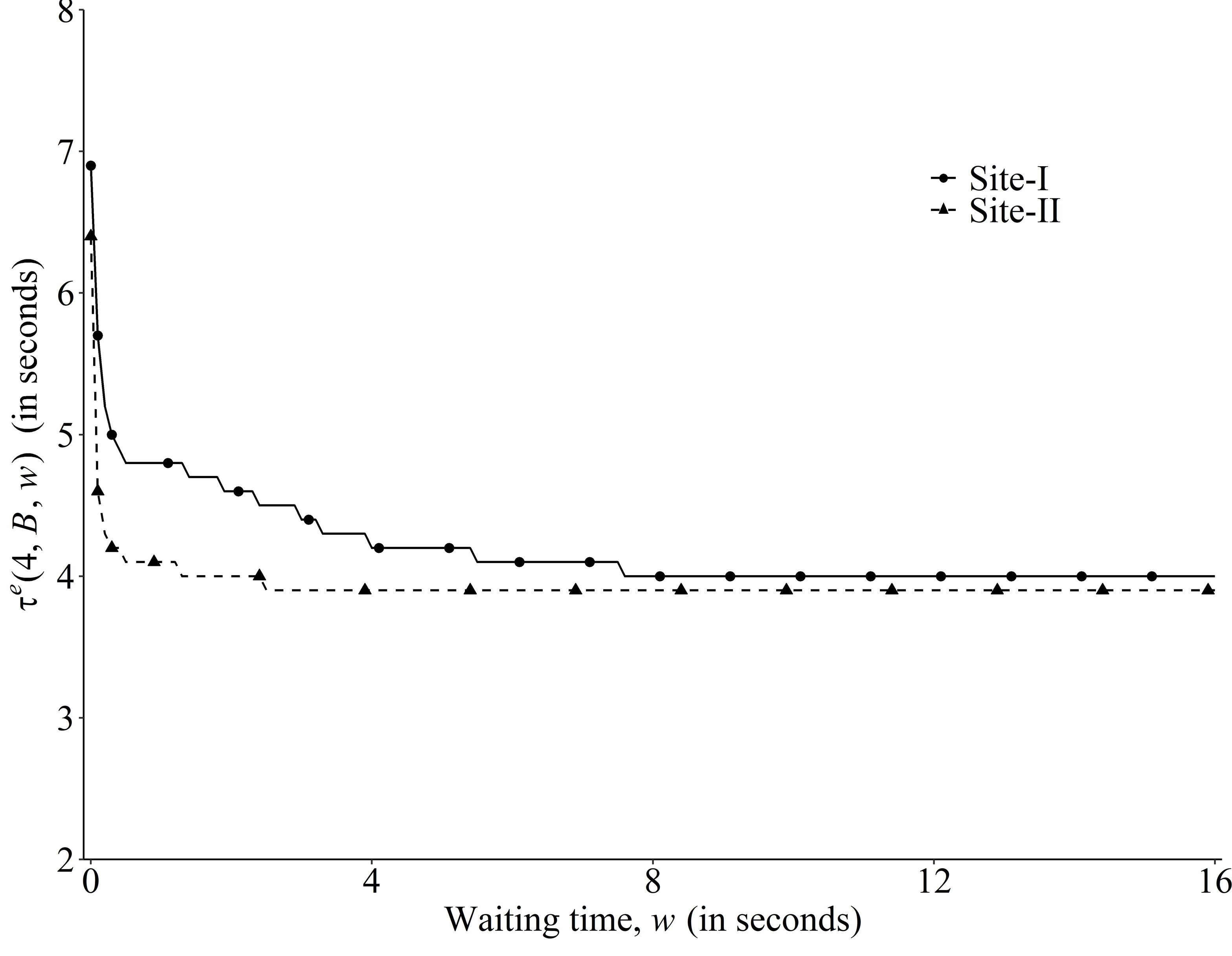}
    \caption{Variation of $\tau^e(4,B)$ with $w$}
    \label{fig:sow_tau_e}
\end{figure}

In the model proposed here, for every combination of $s$ and $o$, the critical gap is a function of $w$, and so is the emulator critical gap. As mentioned in Section \ref{subsubsec_sow}, for every combination of $s$ and $o$, the emulator critical gap, $\tau^e(s,o,w)$ is determined for various values of $w$ using Equation \ref{eq:tcstar_yw}. In order to give an idea about how $\tau^e(s,o,w)$ varies with $w$, $\tau^e(4,B,w)$ versus $w$ graphs are plotted for both the sites in Figure \ref{fig:sow_tau_e}. Plots for other combinations of $s$ and $o$ at each of the sites are similar to those shown in Figure \ref{fig:sow_tau_e} but are not included here to avoid repetition. Two features emerge from these plots: (i) at each site $\tau^e(s,o,0)$ appears to be quite different from the rest of the values of $\tau^e(s,o,w)$, and (ii) as $w$ increases the variation in $\tau^e(s,o,w)$ with $w$ reduces and for $w>8$ seconds (for Site-I) and $w>3.5$ seconds (for Site-II) the emulator critical gap appears to stabilize to a constant value. The $\tau^e(s,o,w)$ values for $w=0$ seconds and large values of $w$ (say, $w=10$ seconds) along with their standard errors for both sites are provided in Table \ref{tab:tau_sow_015}.

\begin{table}[htbp]
    \centering
    \caption{$\tau^e(s,o,w)$ estimates for $w=0$ seconds and $w=10$ seconds}
    \label{tab:tau_sow_015}
    \begin{tabular}{ccclcclcclcc}
        \hline
        \multicolumn{1}{l}{}     & \multicolumn{5}{c}{Site-I} &  & \multicolumn{5}{c}{Site-II}  \\
        \cline{2-6} \cline{8-12}
        \multirow{2}{*}{$s,o$} & \multicolumn{2}{c}{$\tau^e{(s,o,0)}$} &  & \multicolumn{2}{c}{$\tau^e{(s,o,10)}$} &  & \multicolumn{2}{c}{$\tau^e{(s,o,0)}$} &  & \multicolumn{2}{c}{$\tau^e{(s,o,10)}$} \\
        \cline{2-3} \cline{5-6} \cline{8-9} \cline{11-12}
         & Estimate & SE &  & Estimate  & SE  &  & Estimate  & SE  &  & Estimate  & SE \\
         \hline
        2,S   & 4.6  & 0.30   &  & 3.0  & 0.30  &  & 4.2   & 0.15  &  & 2.9   & 0.08     \\
        2,B   & 5.9  & 0.33   &  & 3.6  & 0.15  &  & 5.0   & 0.25  &  & 3.5   & 0.12     \\
        4,S   & 6.1  & 0.26   &  & 2.9  & 0.19  &  & 4.7   & 0.26  &  & 3.6   & 0.16     \\
        4,B   & 6.9  & 0.26   &  & 4.0  & 0.09  &  & 6.4   & 0.53  &  & 3.9   & 0.20     \\
        \hline
    \end{tabular}
\end{table}

Also, for each subject vehicle type and opposing vehicle type, emulator critical gaps are compared at zero waiting time and after it stabilizes. For each site, for $w=0$ seconds, the following hypotheses are tested: \\
$H_0:\tau^e(2,S,0)=\tau^e(2,B,0),\ H_1:\tau^e(2,S,0)<\tau^e(2,B,0)\ $;
$H_0:\tau^e(4,S,0)=\tau^e(4,B,0),\ H_1:\tau^e(4,S,0)<\tau^e(4,B,0)\ $;
$H_0:\tau^e(2,S,0)=\tau^e(4,S,0),\ H_1:\tau^e(2,S,0)<\tau^e(4,S,0)\ $;
$H_0:\tau^e(2,B,0)=\tau^e(4,B,0),\ H_1:\tau^e(2,B,0)<\tau^e(4,B,0)\ $.\\
For Site-I, all the null hypotheses are rejected (at 0.05 level of significance) in favor of the alternate hypotheses. Similar results are obtained for Site-II except that $H_0:\tau^e(2,S,0)=\tau^e(4,S,0)$ was not rejected at 0.05 level of significance. A similar set of hypotheses are tested at each site for $w=10$ seconds. Except for $H_0:\tau^e(2,S,10)=\tau^e(4,S,10)$ at Site-I and $H_0:\tau^e(4,S,10)=\tau^e(4,B,10)$ at Site-II, all other null hypotheses were rejected at 0.05 level of significance. These tests further lend credence to the notion that even at different waiting times critical gap (or its emulator) is smaller when the subject vehicle is more maneuverable, and the critical gap (or its emulator) employed is smaller when a small opposing vehicle approaches.

The sharp reduction in the $\tau^e(s,o,w)$ value may indicate that when drivers do not wait at all before taking a gap (i.e., $w=0$ seconds) they are extra cautious, they accept a gap only when it is large. However, once they stop (i.e., $w>0$ seconds), they tend to accept gaps that are smaller, probably due to the fact that drivers feel more confident in their evaluation of oncoming gaps once they are stopped. This line of thought suggests that the dependence of the critical gap on waiting time can be studied in a simplified manner by developing or estimating two values of critical gap --- one for $w=0$ seconds and the other for $w>0$ seconds. This modified idea of the dependence is implemented by representing $\tau(s,o,0)$ as $\tau_{soz}$ and $\tau(s,o,w>0)$ as $\tau_{son}$. That is, now the critical gap for every $s$ and $o$ combination is bi-valued, one for $w=0$ seconds and the other for $w>0$ seconds. The likelihood function for this estimation is similar to the one shown in Equation \ref{eq:likelihood_tc_s}, except now there are in all eight critical gap parameters. Table \ref{tab:tau_e_sow_discrete} gives the estimates of these parameters. All the estimates are statistically significant. Also, for every $s$ and $o$ combination $\tau^e_{soz}$ is statistically significantly different from $\tau^e_{son}$. However, statistically speaking and as expected, comparisons of AIC values obtained through log-likelihood values in Tables \ref{tab:tau_sow} and \ref{tab:tau_e_sow_discrete} show that the waiting time dependence representation of Equation \ref{eq:tau_sow} (Table \ref{tab:tau_sow}) functions better than its bi-valued approximation (Table \ref{tab:tau_e_sow_discrete}) even after adjusting for the fact that the approximation uses four less parameters. So, while it is easier to implement the bi-valued representation, it is possibly less accurate than the representation in Equation \ref{eq:tau_sow}. Finally, note that the model when $\tau$ depends on $s$ and $o$ is a restricted version of the model when $\tau$ depends on $s$, $o$, and $w$, where waiting time dependence is modeled as a bi-valued approximation. The restrictions are $\tau^e_{2Sz}=\tau^e_{2Sn}=\tau^e(2,S),\ \tau^e_{2Bz}=\tau^e_{2Bn}=\tau^e(2,B),\  \tau^e_{4Sz}=\tau^e_{4Sn}=\tau^e(4,S),\text{ and } \tau^e_{4Bz}=\tau^e_{4Bn}=\tau^e(4,B)$. The likelihood ratio test was performed to compare the likelihood of
these two models given in Table \ref{tab:tau_so} and \ref{tab:tau_e_sow_discrete} for each site. The null hypothesis that restrictions are true is rejected at 0.05 level of significance. This indicates that the inclusion of $w$ as an influencing variable, irrespective of the representation, is an improvement over when $w$ is ignored as an explanatory variable. 

\begin{table}[h]
    \centering
    \caption{Parameters estimates when $\tau$ depends on $s$, $o$ and $w$ (\textbf{discrete case}) }
    \label{tab:tau_e_sow_discrete}
    \begin{tabular}{cccccccc}
        \hline
         \multirow{2}{*}{Parameter} &\multicolumn{3}{c}{Site-I} & & \multicolumn{3}{c}{Site-II} \\
         \cline{2-4} \cline{6-8}
           & Estimate & SE& Max. $LL$ & & Estimate & SE& Max. $LL$     \\
         \hline
           $\alpha/\beta$  & 7.39  &  0.42 & \multirow{11}{*}{-1531.59} & &6.30 &.1.41 &  \multirow{11}{*}{-1317.23} \\  
           $v$  & 0.25  & 0.03 & & & 0.27& 0.03 &   \\
           $k$  & 0.48  & 0.04 & & & 0.32& 0.06   & \\
           $\tau_{2Sz}/\beta$  & 4.57  & 0.32 &  & &4.16& 0.17 &    \\
           $\tau_{2Sn}/\beta$  & 3.32  & 0.20 &  & &2.92&0.08 &    \\
           $\tau_{2Bz}/\beta$  & 5.78  & 0.32 &  & &4.95&0.25 &    \\
           $\tau_{2Bn}/\beta$  & 3.77  & 0.12 &  & &3.51&0.11 &    \\
           $\tau_{4Sz}/\beta$  & 5.84  & 0.26 &  & &4.53&0.31 &    \\
           $\tau_{4Sn}/\beta$  & 3.39  & 0.15 &  & &3.60&0.14 &    \\
           $\tau_{4Bz}/\beta$  & 6.73  & 0.22 &  & &6.33&0.68 &    \\
           $\tau_{4Bn}/\beta$  & 4.26  & 0.11 &  & &3.84&0.16 &    \\
           \hline
           $\tau^e_{2Sz}$  & 4.60  & 0.32 &  & &4.20&0.15 &    \\
           $\tau^e_{2Sn}$  & 3.20  & 0.21 &  & &3.00&0.07 &    \\
           $\tau^e_{2Bz}$  & 5.90  & 0.31 &  & &4.90&0.23 &    \\
           $\tau^e_{2Bn}$  & 3.70  & 0.11 &  & &3.60&0.11 &    \\
           $\tau^e_{4Sz}$  & 6.00  & 0.25 &  & &4.60&0.28 &    \\
           $\tau^e_{4Sn}$  & 3.30  & 0.14 &  & &3.70&0.14 &    \\
           $\tau^e_{4Bz}$  & 6.90  & 0.21 &  & &6.10&0.54 &    \\
           $\tau^e_{4Bn}$  & 4.30  & 0.09 &  & &3.90&0.15 &    \\
         \hline
    \end{tabular}
\end{table}

\subsection{\texorpdfstring{$\tau$}{tau} depends on \texorpdfstring{$s$}{s}, \texorpdfstring{$o$}{o} and \texorpdfstring{$r$}{r}}

\begin{figure} [H]
    \centering
    \includegraphics[width = 0.8 \linewidth]{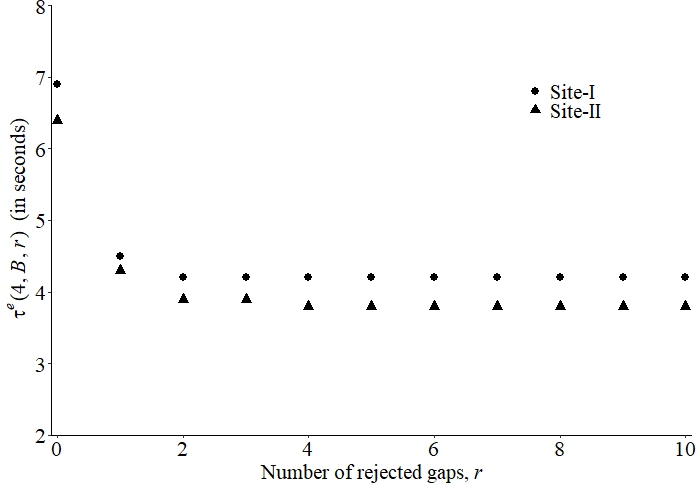}
    \caption{Variation of $\tau^e(4,B)$ with $r$}
    \label{fig:sor_tau_e}
\end{figure}

Critical gap estimation with waiting time dependence is more complicated than the other cases discussed here because one has to incorporate the perceived value of waiting time while modeling its effect on the critical gap. A suggested way around this is to use the number of gaps rejected, $r$, as a surrogate for waiting time. The model for this was presented in Section \ref{subsubsec_sor}, and the parameters are estimated using the likelihood function introduced there. The estimated parameters are shown in Table \ref{tab:tau_sor} and the variation of emulator critical gap, $\tau^e(4,B,r)$, with $r$ is shown in Figure \ref{fig:sor_tau_e}. The variation of the $\tau^e(\cdot)$s for the other $s$ and $o$ combinations with $r$ are similar to those shown in Figure \ref{fig:sor_tau_e} and are not included here to avoid repetition.
\begin{table}[h]
    \centering
    \caption{Parameters estimates when $\tau$ depends on $s$, $o$ and $r$}
    \label{tab:tau_sor}
    \begin{tabular}{cccccccc}
        \hline
         \multirow{2}{*}{Parameter} &\multicolumn{3}{c}{Site-I} & & \multicolumn{3}{c}{Site-II} \\
         \cline{2-4} \cline{6-8}
           & Estimate & SE& Max. $LL$ & & Estimate & SE& Max. $LL$     \\
         \hline
           $\alpha/\beta$  & 7.39  &  0.42 & \multirow{15}{*}{-1528.60} & &6.29 &1.28 &  \multirow{15}{*}{-1316.89} \\  
           $v$  & 0.24  & 0.03 & & & 0.27& 0.03 &   \\
           $k$  & 0.48  & 0.04 & & & 0.33& 0.05   & \\
           $a_{2,S}/\beta$  & 1.39 & 0.44 &  & &1.17&0.18 &    \\
           $c_{2,S}/\beta$  & 3.20  & 0.31 &                    & &2.94&  0.07   &                      \\
           $l_{2,S}      $ & 0.66  & 1.03 &                    & &0.21&  0.20   &                      \\
           $a_{2,B}/\beta$  & 2.02  & 0.34 &                    & &1.52&  0.30   &                       \\
           $c_{2,B}/\beta$  & 3.75  & 0.12 &                    & &3.47&  0.12   &                      \\
           $l_{2,B}      $ & 0.27  &0.20 &                    & &0.43& 0.23   &                      \\
           $a_{4,S}/\beta$  & 2.56  & 0.30 &                    & &1.04&  0.36   &                       \\
           $c_{4,S}/\beta$  & 3.25  & 0.24 &                    & &3.59&  0.16   &                      \\
           $l_{4,S}      $ & 0.54  & 0.57 &                    & &0.07&  0.19   &                      \\
           $a_{4,B}/\beta$  & 2.57  & 0.26 &                    & &2.91& 0.72   &                      \\
           $c_{4,B}/\beta$  & 4.15  &0.13 &                    & &3.75&  0.21   &                      \\
           $l_{4,B}      $ & 0.47  & 0.17 &                    & &0.56&  0.31   &                      \\

         \hline
    \end{tabular}
\end{table}

It is observed from these plots that for each site, $\tau^e(s,o,0)$ is quite different from $\tau^e(s,o,r)$ when $r \neq 0$. Also, as the number of rejected gaps, $r$, increases, $\tau^e(s,o,r)$ decreases at a reducing rate and stabilizes to a constant value for $r>3$. These observations are in line with how $\tau^e(s,o,w)$ behaves with changes in $w$. The $\tau^e(s,o,r)$ values for $r=0$ and for large values of $r$ (say, $r=4$) along with their standard errors for both sites are provided in Table \ref{tab:tau_sor_06}.

\begin{table}[h]
    \centering
    \caption{$\tau^e(s,o,r)$ estimates for $r=0$ and $r=4$}
    \label{tab:tau_sor_06}
    \begin{tabular}{ccclcclcclcc}
        \hline
        \multicolumn{1}{l}{}     & \multicolumn{5}{c}{Site-I} &  & \multicolumn{5}{c}{Site-II}  \\
        \cline{2-6} \cline{8-12}
        \multirow{2}{*}{$s,o$} & \multicolumn{2}{c}{$\tau^e{(s,o,0)}$} &  & \multicolumn{2}{c}{$\tau^e{(s,o,4)}$} &  & \multicolumn{2}{c}{$\tau^e{(s,o,0)}$} &  & \multicolumn{2}{c}{$\tau^e{(s,o,4)}$} \\
        \cline{2-3} \cline{5-6} \cline{8-9} \cline{11-12}
         & Estimate & SE &  & Estimate  & SE  &  & Estimate  & SE  &  & Estimate  & SE \\
         \hline
        2,S   & 4.6  & 0.34   &  & 3.1  & 0.25  &  & 4.2  & 0.15   &  & 3.0  & 0.07    \\
        2,B   & 5.9  & 0.33   &  & 3.7  & 0.13  &  & 5.0  & 0.24   &  & 3.6  & 0.12     \\
        4,S   & 6.0  & 0.32   &  & 3.1  & 0.18  &  & 4.7  & 0.28   &  & 3.7  & 0.15     \\
        4,B   & 6.9  & 0.22   &  & 4.2  & 0.14  &  & 6.4  & 0.56   &  & 3.8  & 0.19     \\
        \hline
    \end{tabular}
\end{table}

Emulator critical gaps, $\tau^e(s,o,r)$, are compared for each subject vehicle type and opposing vehicle type combination when the number of rejected gaps is zero $(r=0)$ and after it stabilizes $(r=4)$. For each site, for $r=0$, the following hypotheses are tested: \\
$H_0:\tau^e(2,S,0)=\tau^e(2,B,0),\ H_1:\tau^e(2,S,0)<\tau^e(2,B,0)$\ ; $H_0:\tau^e(4,S,0)=\tau^e(4,B,0),\ H_1:\tau^e(4,S,0)<\tau^e(4,B,0)$\ ;
$H_0:\tau^e(2,S,0)=\tau^e(4,S,0),\ H_1:\tau^e(2,S,0)<\tau^e(4,S,0)$\ ;
$H_0:\tau^e(2,B,0)=\tau^e(4,B,0),\ H_1:\tau^e(2,B,0)<\tau^e(4,B,0)$.\\
For Site-I, all the null hypotheses are rejected (at 0.05 level of significance) in favor of the alternate hypotheses. Similar results are obtained for Site-II except that $H_0:\tau^e(2,S,0)=\tau^e(4,S,0)$ was not rejected at 0.05 level of significance. A similar set of hypotheses are tested at each site for $r=4$. Except for $H_0:\tau^e(2,S,4)=\tau^e(4,S,4)$ at Site-I and $H_0:\tau^e(4,S,4)=\tau^e(4,B,4)$ and $H_0:\tau^e(2,B,4)=\tau^e(4,B,4)$ at Site-II, all other null hypotheses were rejected at 0.05 level of significance. These results again suggest that for different number of rejected gaps, the emulator critical gap is smaller when the subject vehicle is a two-wheeler, and also when the vehicle at the end of the gap is small. 

$r$ is suggested as a surrogate for $w$ since, in general, as $r$ increases so should $w$. Of course, given that gap sizes can be small or large one cannot develop a deterministic relation between $w$ and $r$. However, $r=0$ and $w=0$ refer to the same situation as do the values of $r$ and $w$ where the critical gap stabilizes. So, it is worthwhile to compare the values of $\tau^e(s,o,w)$ and $\tau^e(s,o,r)$ when (i) both $r$ and $w$ are zero and (ii) both $r$ and $w$ are large, say $r=4$ and $w=10$. The values of $\tau^e(s,o,w)$ to be used in the comparison are in Table \ref{tab:tau_sow_015} and those for $\tau^e(s,o,r)$ are in Table \ref{tab:tau_sor_06}. The hypotheses tested in this case are $H_0:\tau^e(s,o,r)=\tau^e(s,o,w),\ H_1:\tau^e(s,o,r)=\tau^e(s,o,w)$, first when $r$ and $w$ are both zeros and then when $r=4$ and $w=10$. In neither case, the null hypothesis could be rejected at 0.05 level of significance.

The observations from the figure, and the comparisons of $\tau^e(\cdot)$ values suggest that the number of rejected gaps, $r$, is a good surrogate for waiting time, $w$, and can be used for real-world applications. Also, it is easier to collect data on $r$ than $w$. AIC comparisons, however, show that $\tau^e(s,o,w)$ explains more than the $\tau^e(s,o,r)$ version.\footnote{It may be noted though that log-likelihood ratio test between the $\tau(s,o)$ and $\tau(s,o,r)$ models show that the latter model is an improvement over the former.} That is, while the model with \textquotedblleft$r$\textquotedblright\ is simpler and data on \textquotedblleft$r$\textquotedblright\ is easier to collect the model with \textquotedblleft$w$\textquotedblright\ performs better. It is for this reason that \textquotedblleft$r$\textquotedblright\ is introduced as a surrogate for \textquotedblleft$w$\textquotedblright.

Before leaving the section on parameter estimates, it is worthwhile to mention that in order to study whether the parameter estimates are retrievable, synthetic data are generated using assumed parameter values for (i) the simplest case (when $\tau$ is independent of external influence) and (ii) the most complex case (when $\tau$ depends on $s,\ o$ and $w$). These studies indicate that the parameters are retrievable --- the estimated parameters become indistinguishable from the assumed parameters for synthetic datasets of size 5000 or larger. 

\section{Conclusion} \label{sec:conclusion}
The present work attempts to estimate the critical gap (which, it is argued, is a latent variable) while explicitly modeling that decision-making on acceptance/rejection of gaps happens in a person's mind using the perceived value of gaps and not the measured or observed values. It is also argued that variability in critical gaps can, at least to some extent, be attributed to variations in certain exogenous variables (referred to as influencing variables) like subject and opposing vehicle types, waiting time, etc. 

Various models of the gap acceptance process under different assumptions on its variability and influencing factors are presented. Likelihood functions for each of these models are also developed. Maximum likelihood estimates of the model parameters using data from two sites do lend credence to the assumptions related to the impact of influencing variables. For example, when subject type, $s$, is assumed to influence the critical gap, the results clearly show that the addition of $s$ improves the model predictions on acceptance/rejection. The inclusion of $s$ and opposing vehicle type, $o$, leads to further improvements. Finally, when the critical gap $\tau$ is assumed to be a function of $s,\ o$, and $w$, the gap acceptance formulation performs the best.

The realization that the critical gap is a latent variable implies that one needs to find an emulator critical gap that can work with observed gaps to provide answers as to which gap will be accepted and which will be rejected. Such an emulator critical gap is also developed and estimated under the various assumptions of influencing variables. It is satisfying to see that the use of the emulator critical gaps estimated here predicts observed average waiting times more closely than other methods.

From data on gap sizes and acceptance/rejection, with the help of the proposed models, considerable insights into the gap acceptance behavior were obtained. It was heartening to see that the data clearly corroborated the various hypotheses on the effect of the subject vehicle and opposing vehicle types and waiting time on the critical gap and its emulator. Another interesting result was to establish that the number of rejected gaps can function as a reasonable surrogate for waiting time.

While this work explicitly models the randomness associated with perception and introduces mechanisms to model changes in critical gap due to various influencing variables, it continues to assume that the critical gap is a deterministic variable. The authors have also relaxed this assumption and studied critical gaps when they are modeled as stochastic variables under a similar perceptual framework as here. Various influencing variables are also included in these models. The stochastic variant of the present work is the subject matter of another paper.
\\

\appendix
\renewcommand\theequation{\Alph{section}.\arabic{equation}}
\renewcommand\thetable{\Alph{section}.\arabic{table}}
\renewcommand\thefigure{\Alph{section}.\arabic{figure}}

\section{Derivation of computed average (expected) waiting time, C-AWT} \label{app:exp_wt}

\setcounter{equation}{0}
\setcounter{figure}{0}
\setcounter{table}{0}

Let $W$ be the waiting time before accepting any gap and $G_1, G_2, \cdots, G_m$ be the $1^{st}, 2^{nd}, \cdots, m^{th}$ gaps. Note that these $G_i$ are independent and identically distributed with density function $f_G(g)$.
The expected waiting time, $E[W]$ or C-AWT can be written as,
\begin{align*}   
    E[W]   &= \sum_{m=0}^{\infty} E[W|m^{th}\text{gap accepted}].P(m^{th}\text{gap accepted}) \nonumber \\
           &= E[W|1^{st}\text{gap accepted}].P(1^{st}\text{gap accepted}) + 
           E[W|2^{nd}\text{gap accepted}].P(2^{nd}\text{gap accepted}) + \nonumber \\
           & \ \ \ \  E[W|3^{rd}\text{gap accepted}].P(3^{rd}\text{gap accepted}) +
           E[W|4^{th}\text{gap accepted}].P(4^{th}\text{gap accepted}) 
           \cdots 
\end{align*}

Let $p_o$ be the probability that a gap $G_i$ is accepted; i.e., $p_o = P(G_i > \tau^e) = 1-F_G(\tau^e)$. Therefore,
\begin{align} 
    E[W]   &= (0\times p_o) + \left(E[G_1|G_1<\tau^e]\times (1-p_o)p_o\right)  + \nonumber \\ 
           & \ \ \ \  \left((E[G_1+G_2|G_1<\tau^e,G_2<\tau^e])\times (1-p_o)^2 p_o\right) + \nonumber \\
           & \ \ \ \  \left((E[G_1+G_2+G_3|G_1<\tau^e,G_2<\tau^e,G_3<\tau^e])\times (1-p_o)^3 p_o\right) + \cdots \nonumber\\ 
           &= 0 + \left(E[G_1|G_1<\tau^e])\times (1-p_o)p_o\right)  + \nonumber \\
           & \ \ \ \  \left((E[G_1|G_1<\tau^e,G_2<\tau^e]+E[G_2|G_1<\tau^e,G_2<\tau^e])\times (1-p_o)^2 p_o\right) + \nonumber \\
           & \ \ \ \  ((E[G_1|G_1<\tau^e,G_2<\tau^e,G_3<\tau^e]+E[G_2|G_1<\tau^e,G_2<\tau^e,G_3<\tau^e]+ \nonumber \\ 
           & \ \ \ \  E[G_3|G_1<\tau^e,G_2<\tau^e,G_3<\tau^e])\times (1-p_o)^3 p_o) + \cdots \nonumber
\end{align}
Since $G_1, G_2, \cdots, G_m$ are independent, therefore, for $k \not = j$ and $\forall k,j \in (1,2,\cdots,m)$,\\ $E[G_j|G_j<\tau^e,G_k<\tau^e, \cdots] = E[G_j|G_j<\tau^e]$. So,
\begin{align}  
    E[W]   &= \left(E[G_1|G_1<\tau^e]\times (1-p_o)p_o\right)  + \left(E[G_1|G_1<\tau^e]+E[G_2|G_2<\tau^e])\times (1-p_o)^2 p_o\right) + \nonumber \\
           & \ \ \ \  \left(E[G_1|G_1<\tau^e]+E[G_2|G_2<\tau^e]+ E[G_3|G_3<\tau^e])\times (1-p_o)^3 p_o\right) + \cdots \nonumber
\end{align}

Also, as $G_1, G_2, \cdots, G_m$ are identical, therefore, $E[G_j|G_j<\tau^e] = E[G|G<\tau^e] \ \  \forall j \in (1,2,\cdots,m)$. 
\begin{align} \label{eq:E_wt}  
    E[W]   &= \left(E[G|G<\tau^e]\times (1-p_o)p_o\right)  \left((E[G|G<\tau^e]+E[G|G<\tau^e])\times (1-p_o)^2 p_o\right) + \nonumber \\
           & \ \ \ \  \left((E[G|G<\tau^e]+E[G|G<\tau^e]+ E[G|G<\tau^e])\times (1-p_o)^3 p_o \right) + \cdots \nonumber\\
           &= \left(E[G|G<\tau^e]\times (1-p_o)p_o\right)  + \left(2 E[G|G<\tau^e]\times (1-p_o)^2 p_o\right)  + \nonumber \\
           & \ \ \ \  \left(3 E[G|G<\tau^e]\times (1-p_o)^3 p_o\right) + \cdots \nonumber \\
           &= E[G|G<\tau^e]\times (1-p_o)p_o \left(1+2(1-p_o)+3(1-p_o)^2+\cdots \right) \nonumber \\
           &= \frac{\int_{g=0}^{\tau^e} g\ f_G(g)dg}{P(G<\tau^e)}  \times (1-p_o)p_o \times \frac{1}{p_o^2}\nonumber \\ 
           &= \frac{\int_{g=0}^{\tau^e} g\ f_G(g)dg}{1-p_o}  \times (1-p_o) \times \frac{1}{p_o} \nonumber \\
           &= \frac{\int_{g=0}^{\tau^e} g\ f_G(g)dg}{1-F_G(\tau^e)} 
\end{align}

Therefore,
\begin{equation} \label{eq:CAWT}
    \text{C-AWT}=\frac{\int_{g=0}^{\tau^e} g\ f_G(g)dg}{1-F_G(\tau^e)}
\end{equation}

C-AWT for subject vehicle type, $s$, can be found using $\tau^e(s)$ in place of $\tau^e$ in Equation \ref{eq:CAWT}.

\bibliography{ref}

\end{document}